# Does a Self-Learning Migration Agent Learn? An Audited Evaluation of Evidence-Graded Experience Memory for Oracle-to-PostgreSQL Migration

Oleg Grynets
EPAM Systems
McLean, Virginia, USA
oleg_grynets@epam.com

Vasyl Lyashkevych
EPAM Systems
Lviv, Ukraine
vasyl_lyashkevych@epam.com

Valeriia Oliferuk
EPAM Systems
Vinnytsia, Ukraine
valeriia_oliferuk@epam.com

Vladyslav Babenko
EPAM Systems
Odesa, Ukraine
vladyslav_babenko1@epam.com

Danylo Kovalchuk
EPAM Systems
Lviv, Ukraine
danylo_kovalchuk@epam.com

Dmytro Kostetskyi
EPAM Systems
Lviv, Ukraine
dmytro_kostetskyi@epam.com

***Abstract*—It is difficult to verify the claim that an agent based on a large language model (LLM) learns from its own experience. Performance improvements could stem from information retrieval, an increased number of attempts, or the use of reference target data accessed by the agent via its own memory. Therefore, this paper presents an audit of the evaluation process and the results of an investigation into a self-learning agent, designed for migrating databases from Oracle to PostgreSQL, that includes an assessment of the reliability of the evidence. The agent retains experience, adopts rules based on confidence-level weighting, and adapts its strategy without updating the weight coefficients. On 1,833 documentation-derived Oracle–PostgreSQL pairs (367 held-out units in 136 groups, Qwen3-Coder-480B), persistent memory raised reference conformity (RC) over a no-memory baseline by 0.015 [0.004, 0.025], and the adaptive policy added nothing (−0.004 [−0.010, 0.001]). The audit found that 72% of held-out units share their PostgreSQL reference with a training pair. On the 103 leak-free units the memory effect fell to 0.002 [−0.008, 0.011]. A model-free baseline that copies the target of the most similar training source reached RC 0.222 and beat every LLM system (at most 0.163). On a new split grouped by reference, copying falls to 0.119, and a rerun on 374 test units in 118 reference groups found no memory effect at group level (0.005 [−0.003, 0.017]) and no policy effect (0.000); 88% of the unit-level gain came from one reference shared by 42 test units. A study based on evaluating the execution of 216–224 real-world PL/SQL units using Oracle data revealed that PostgreSQL's acceptance mechanism overestimates correctness (10–22% of accepted units were incorrect) and that an agent that does nothing performs well on metrics that reward silence. Employing a pre-registered, human-supervised correction loop increased the number of clearly correct results from 98 out of 216 to 167 out of 224. An analysis of the knowledge base update process (write-back) revealed that increasing the knowledge base size from 12 to 1,200 entries reduces retrieval reliability margins by 54%. Based on these findings, we formulate six rules for evaluating claims regarding the learning capabilities of agents.**

## I. Introduction

### *A. Background and Motivation*

Agents based on LLMs are increasingly being used as tools for software development tasks that require consideration of repository context and iterative feedback [1]–[3]. Code translation and database migration are natural targets for such agents, because migration units are numerous, structurally repetitive, and linked by dependencies, yet they also contain dialect-specific behavior whose semantics fluent output can silently distort [4]–[6].

Oracle-to-PostgreSQL migration illustrates the difficulty. Oracle SQL and PL/SQL combine constructs such as `NVL`, `DECODE`, `CONNECT BY`, `ROWNUM`, `MERGE`, package state, autonomous transactions, and `DBMS_*` calls, each of which may require a context-dependent PostgreSQL idiom. Earlier work by the authors' group examined a fine-tuned migration framework [7] and token-optimization strategies for Oracle-to-PostgreSQL migration [8]. Related work also addresses specification-driven synthesis for intelligent monitoring systems [9].

Feedback-driven agents improve outputs through reflection, self-refinement, execution feedback, or self-debugging [10]–[12]. In most of these methods, however, the improvement is task-local: the same task is attempted repeatedly, and the benefit disappears when the next task starts. Whether an agent becomes better on *previously unseen* migration tasks because of experience accumulated on other tasks is a distinct question, and it is rarely separated from retrieval of similar examples, copying of references, or simply spending more inference budget [13], [14].

Additional motivation comes from earlier research on knowledge acquisition. Situation-oriented ontologies organized reusable diagnostic knowledge based on context and relationships, while tools for working with tabular knowledge bases supported the process of their creation [15], [16]. In cases of migration, we apply only the principles of situational suitability and clearly defined provenance. These materials are not presented as a demonstration of agent training based on LLMs.

Our initial plan was to demonstrate that such an agent learns. The plan changed as soon as the results came in. Memory did help, but the gains were almost entirely concentrated on stored items that were also referenced during training. A program that simply copies the nearest training target performed better than any LLM-based system. Thus, this article describes an agent, an audit that uncovered a problem, the remediation of a gap, and two laboratory studies demonstrating which specific aspects can be safely optimized through a learning loop. The agent remains a subject of scientific research, with the migration from Oracle to PostgreSQL serving as a test environment.

### *B. Research Questions*

The core question is whether continuous, evidence-based learning from previous migration pairs provides an advantage on unseen tasks compared to static prompting, nearest-example retrieval, and task-specific local reflection, under the same base LLM and inference budget. We have broken this down into four research questions, one for each study:

- RQ1 (Study 1): Do experience (memory) and an adaptive policy strategy improve alignment with reference patterns (on data not used for training) compared to baselines, and does each of these components contribute to this result?
- RQ2 (Study 2): What proportion of the observed uplift remains after accounting for shared links and comparing against baseline copying metrics derived without the use of models—and how does the picture change when the data is grouped by reference target?
- RQ3 (Study 3): When results are evaluated by executing them and comparing them against Oracle execution ground truth, what specific signals can the training loop rely on for optimization—without being misled? And does a human-supervised correction loop actually provide real value?
- RQ4 (Study 4): Does the knowledge write-back mechanism remain reliable as the knowledge base grows?

Studies 1 and 2 utilize the same dataset and the original execution protocols from Study 1. Additionally, Study 2 involves a separate re-run of the agent on a sample generated using a reference-based grouping approach. Studies 3 and 4 are part of other laboratory projects and rely on their own distinct datasets. In each study, the answer to the research question is formulated within the framework of the respective approach, and these answers are synthesized in Section XI.

### C. Scope and Contributions

Studies 1 and 2 utilize a single dataset derived from documentation. Since no executable Oracle instance is available for this dataset, the outcome is measured by alignment with reference data rather than by the correctness of behavior. Study 3 employs different toolkits that use actual Oracle output and a PostgreSQL database for verification. Study 4 is a controlled information retrieval experiment. The base model remains unchanged; we therefore characterize the mechanism as non-parametric continual adaptation.

The article makes five contributions. First, it specifies an evidence-graded learning agent with tiers G0–G3, verifier-grounded admission and a checkpointed policy (Sections III–V). Second, it reports a full run of that agent with ablations and controls, including a null result for policy learning (Section VII). Third, it gives a leakage audit with model-free copy baselines and a checkpoint monitor, shows that the memory gain is mostly reference reuse, builds a split grouped by reference and reruns the agent on it (Section VIII). Fourth, it reports execution-judged evidence on what accepted, empty and judged outputs really mean (Section IX) and on write-back at scale (Section X). Fifth, it turns these results into rules for claims of agent learning (Section XI).

## II. Related Work

### A. Feedback-Driven and Self-Improving Code Agents

Reflexion stores verbal reflections in an episodic buffer to improve later attempts, and Self-Refine iterates on a single output using the model's own critique [10], [11]. CodeRL and LEVER use execution outcomes to train or rerank code generation, and Self-Debug and LLMLOOP feed execution or test feedback back into repair loops [12], [17], [18]. LLMLOOP in particular packages iterative compilation, static analysis, and test feedback for generated code and tests [19]. These methods provide the task-local reflection baseline used here, but they do not by themselves establish that experience transfers across tasks.

DSPy compiles declarative language-model pipelines and optimizes prompts or demonstrations against a metric, which makes it a relevant comparator for automatically optimized demonstrations [14]. Early-experience agent learning studies how agents can learn from self-generated interaction data without explicit reward signals [13]. The present study differs in that experience units are typed, evidence-graded, and admitted only through deterministic verifier outputs, and in that learning is measured on a frozen held-out panel.

ExpeL collects experiences across tasks and distils them into insights [20], and Voyager keeps a growing skill library [21]. Agentic context engineering treats a playbook as a context that evolves through generation, reflection and curation, without weight updates [22]. These systems are closest to our design, and they share its main risk: the stored experience may simply encode answers to the evaluation tasks.

### B. Agentic Software Engineering and Evaluation

SWE-agent shows that the design of the agent–computer interface strongly affects software-engineering performance, and SWE-bench established realistic repository-level evaluation [1], [2]. Analyses of LLM-based software-engineering agents argue for simple, well-understood baselines before complex agent architectures [3], and RepairAgent, MetaGPT, and SOEN-101 illustrate autonomous repair, role-based multi-agent collaboration, and process-model-driven generation [23]–[25]. For validity, EvalPlus shows that limited test suites can overstate the correctness of generated code, SWT-Bench evaluates whether generated tests validate real-world fixes, and a study on the reliability and explainability of program-generation models documents related evaluation limitations [26]–[28].

Leakage between training and evaluation data is a known cause of over-optimistic results in machine-learning research [29]. LLM judges such as G-Eval help score open-ended output [30], but their stability on repeated calls is rarely reported. For migration, a hybrid decomposition framework from our group measured routing, repair and decomposition on 1,448 tasks and found that small samples can reverse a comparison [31]. A metric-based model of software functional state combines several metrics into one index over the development life cycle and adds non-compensatory rules, so that strong secondary metrics cannot hide a critical defect; it was validated on 15 merged open datasets and five release scenarios [32]. RC in Study 1 is a compensatory composite of this kind, which is one reason why Studies 2 and 3 examine its parts separately.

### C. Specification Reconstruction, Translation, and Migration

Explain-then-translate shows that self-generated explanations can improve program translation, which motivates an intermediate textual specification [33]. UniTrans combines test-case generation, translation, and iterative repair for code translation, and XCodeEval provides an execution-based multilingual benchmark that includes translation [4], [5]. Work on repairing ambiguous problem descriptions highlights that underspecified inputs degrade generation, which parallels the risk of inventing business semantics during specification extraction [34]. A knowledge-graph-based framework for LLM-powered database migration addresses system evolution, although its relevance to Oracle-to-PostgreSQL relational migration must be assessed separately [6]. Neuro-symbolic feedback, in which symbolic checks constrain neural repair, is explored for vulnerability repair [35].

### D. Knowledge Representation and Contextual Applicability

Situation-oriented ontology construction and knowledge-base authoring software represent knowledge as contextual cases with typed relations [15], [16]. This is related to, but distinct from, neural parameter learning. In the present design every case stores positive and negative evidence, source and target provenance, a dependency neighborhood, applicability predicates, and the strongest verified evidence tier, and the memory can retract, supersede, and deactivate conflicting cases instead of accumulating every generated explanation.

The presentation [36] introduces self-learning monitoring agents that maintain shared knowledge about evolving software and reports repository-level experiments. Its evidence is treated here as preliminary, not as independent validation of migration-agent learning.

### E. Gap Addressed

Existing work offers task-local feedback loops, experiential agents, benchmarks and translation pipelines. We are not aware of a migration study that grades source–target pairs by evidence tier, measures cross-task gain against a fixed-policy memory control on a frozen panel, and then audits that gain for reference sharing and against model-free copying. We also report what happens when the same question is asked with execution against Oracle.

## III. Problem Definition and Formal Model

### A. Agent State and Learning Criterion

Let $D$ denote the set of Oracle migration units, each with source $s$. A candidate pair is a tuple $p = (s, t, e)$, where $t$ is the aligned PostgreSQL artifact and $e \in \{G0, G1, G2, G3\}$ is its evidence tier. Pairs are partitioned into group-disjoint training, validation, and test sets. The agent state after training batch $k$ is $S_k = (M_k, \pi_k)$, where $M_k$ is the experience memory and $\pi_k$ is the strategy policy. State evolves only through training batches $B_k$:

$$S_k = U\left(S_{k-1}, B_k\right), \quad S_k = \left(M_k, \pi_k\right) \tag{1}$$

The experience store comprises verified reusable knowledge $K_k$ and episodic execution memory. For a training unit $T_i$ with candidate $\hat{P}_i$ and verifier feedback $F_i$ (parser diagnostics, reference-alignment scores, dependency-preservation checks, and optional target-side execution outcomes), knowledge is updated by

$$K_{k+1} = U\left(K_k, Admit\left(Extract\left(T_i, \hat{P}_i, F_i\right)\right)\right) \tag{2}$$

where $U$ applies observation, verification, admission, and policy update. Table I lists exactly what is allowed to change over time; the base model, prompts, temperature, tool rights, and token budgets are held constant.

TABLE I. Components of Agent State That May Change Across Checkpoints

| Component | Role and update rule |
|---|---|
| Experience store $M_k$ | Typed experiences with tier, applicability conditions, contraindications, and negative examples; changed only by admission, revision, or deprecation. |
| Strategy policy $\pi_k$ | Ranker over direct, spec-first, example-assisted, and dependency-decomposed strategies; trained on training and validation outcomes only. |
| Rule confidence | Evidence-weighted estimate per rule, recomputed after each batch (Eq. (10)). |
| Retrieval weighting | Weights of lexical, AST, and graph signals; fixed in B5, adaptive in P1. |
| Base-model weights | Not changed. No parametric training is claimed. |

Everything outside Table I stays fixed between checkpoints, so a change in held-out scores can only come from the rows of the table.

An agent is said to *learn* on the test panel $T$ only if three conditions hold jointly: reference conformity on $T$ increases with checkpoint $k$ under constant model and budget; the gain remains after replacing learned memory with random, shuffled, or unrelated memory; and matched with-versus-without-memory counterfactuals show positive rule utility on eligible tasks. Retrieval of similar examples, prompt refinement on the same task, reference copying, and repeated attempts do not satisfy this definition.

Study 2 adds a fourth condition. The gain must remain on held-out units whose reference never occurs as a training target, and the agent must beat a program that copies the target of the most similar training source. Without this condition an agent can meet the first three conditions by storing near-copies of shared references.

### B. Metrics

Metrics are predeclared and measured only against reference checks and structural verifiers. *Reference conformity* is

$$RC = \frac{N_{pass}}{N_{elig}} \tag{3}$$

where $N_{pass}$ is the number of predeclared reference checks passed and $N_{elig}$ the number of checks eligible for the unit; units without a label are excluded rather than counted as failures. Dependency-preservation rate (DPR) and its coverage are

$$DPR = \frac{|E_{pres}|}{|E_{map}|} \tag{4}$$

$$DPR_{cov} = \frac{|E_{map}|}{|E_{src}|} \tag{5}$$

where $E_{src}$ is the set of source dependency edges, $E_{map}$ the mapped and comparable subset, and $E_{pres}$ the preserved subset of $E_{map}$. Learning gain, negative transfer, token efficiency, and rule utility are

$$LG(k) = Q\left(P1_k, T\right) - Q\left(B5_0, T\right) \tag{6}$$

$$NTR(k) = \frac{1}{N}\sum_{i=1}^{N} 1\left[y_i^{base} = 1 \land y_i^{P1_k} = 0\right] \quad (7)$$

$$TPS = \frac{\sum_i \left(n_i^{in} + n_i^{out}\right)}{N_{pass}} \quad (8)$$

$$RU(r) = \frac{1}{|T_r|}\sum_{i \in T_r}\left[q_i(with\, r) - q_i(without\, r)\right] \quad (9)$$

where $Q$ is mean reference conformity on $T$, $y_i$ is binary success of task $i$, $n_i^{in}$ and $n_i^{out}$ are input and output tokens, and $T_r$ are unseen tasks eligible for rule $r$. Cost in currency is reported only if the price is known; all metrics are disaggregated by object type, complexity, rarity, dependency depth, and evidence tier.

In Study 1 the eligible checks for (3) are four weighted checks: exact match (RC = 1), normalised match (two checks), token-sequence similarity (one check above 0.9, half a check above 0.7) and coverage of key SQL constructs. A unit counts as a binary success when it passes the static dialect check and RC ≥ 0.5. The dialect check is a set of regular expressions: no Oracle-only token may survive and a PostgreSQL marker must be present. It is a proxy, not a parser. DPR (4)–(5) was not computed, because no code-level dependency edges were extracted.

### C. Architecture and Train/Test Boundary

Fig. 1 shows the strict boundary between the adaptation zone and the frozen evaluation zone. Memory and policy snapshots cross the boundary read-only; held-out outputs are verified and scored once and are never written back into memory, policy, or hyperparameters.

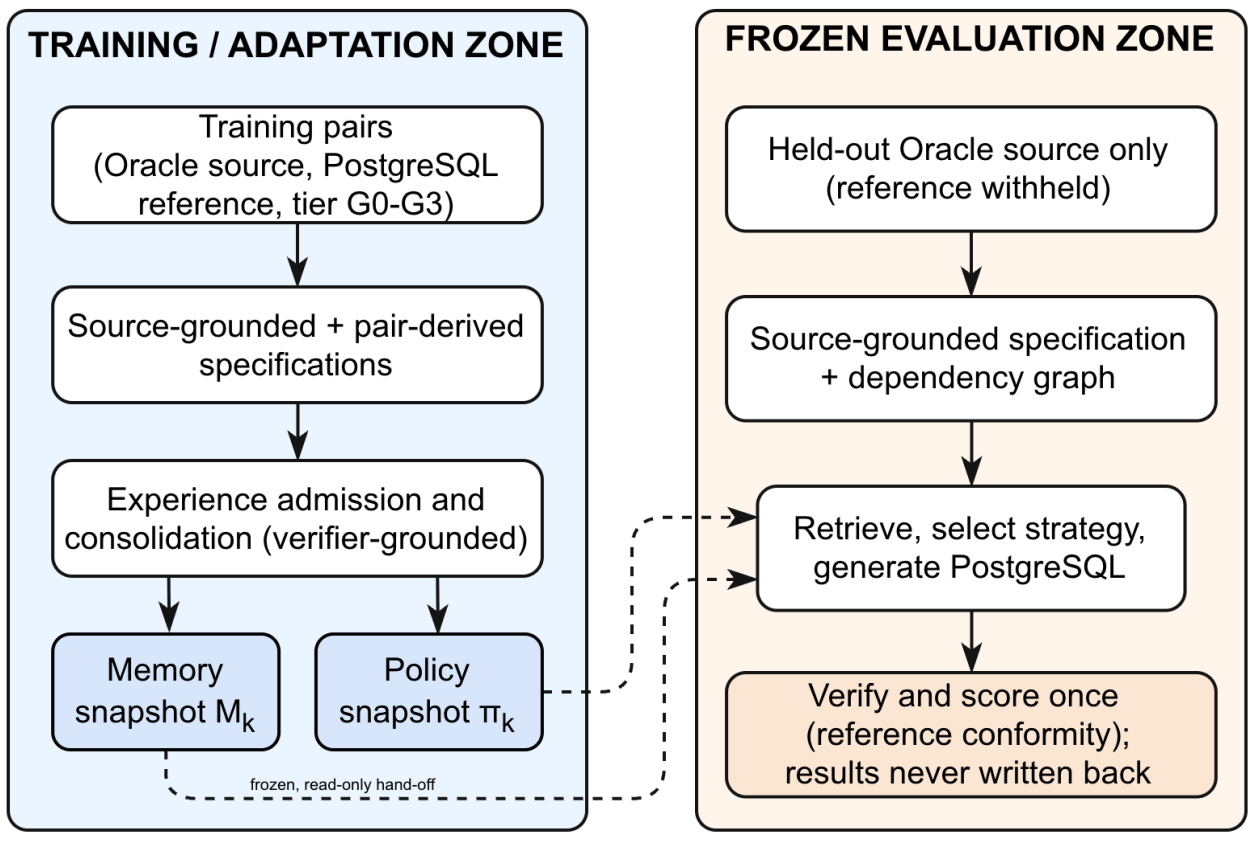

Fig. 1. Architecture with strict train/test boundary. Snapshots are frozen at each checkpoint before held-out scoring.

The boundary in Fig. 1 was enforced by tests on pair and source identifiers. Section VIII shows that this is not enough: a reference can cross the boundary when several sources are aligned to the same target.

## IV. Corpus, Alignment and Evidence Tiers

### A. Evidence Boundaries

Table II states what each study could measure. Earlier migration studies by our group [7], [8] serve as background only; their outcomes are not reused here.

TABLE II. Evidence Matrix

| Item | Status | Consequence |
|---|---|---|
| Oracle sources (Studies 1–2) | 1,834 snippets from the Oracle Database 12.2 SQL and PL/SQL references | Prompts, inventory, copy baselines |
| PostgreSQL references (Studies 1–2) | 9,041 snippets from the PostgreSQL documentation; topical, not translations | RC measures agreement with a related example |
| Code-level dependency graph | Not extracted; only pair links | DPR not computed; H3 not testable |
| Stage A specification audit | Not measured: the audit emitted fixed scores | No specification audit reported |
| Executable Oracle (Study 3) | Oracle transcripts and an Oracle container for 34 files | Execution-judged correctness |
| PostgreSQL deployment | Study 3 only | G2-type evidence outside Study 1 |
| Behavioural equivalence (Study 1) | Not measurable | Reported as such |
| Tokens, calls, latency | Measured from run logs | Table XI |

The most important row is the second one. Studies 1 and 2 measure how close an output comes to a related documentation example, so every result there is a statement about agreement, never about correct migration.

### B. Corpus Discovery and Characterization

The Study 1 corpus comes from two exports of the Oracle documentation and one export of the PostgreSQL documentation. The inventory hashes every snippet, removes exact duplicates and records lines, tokens and detected constructs (Table III). Most raw records are repeats: 14,450 Oracle records reduce to 1,834 unique units and 156,465 PostgreSQL records to 9,041.

TABLE III. Corpus Inventory and Construct Profile

| Quantity | Oracle | PostgreSQL |
|---|---|---|
| Documentation exports | 2 | 1 |
| Raw records → unique units | 14,450 → 1,834 | 156,465 → 9,041 |
| Lines of code / tokens | 12,279 / 99,629 | 83,725 / 745,745 |
| Rare constructs (< 5 units) | DECODE (2), MERGE (2) | none |
| Code-level dependency edges | not extracted | not extracted |

Fig. 2 shows the construct profile. Oracle units are dominated by DBMS_* packages, DUAL and SYSDATE; PostgreSQL units by PL/pgSQL blocks, cursors and generate_series.

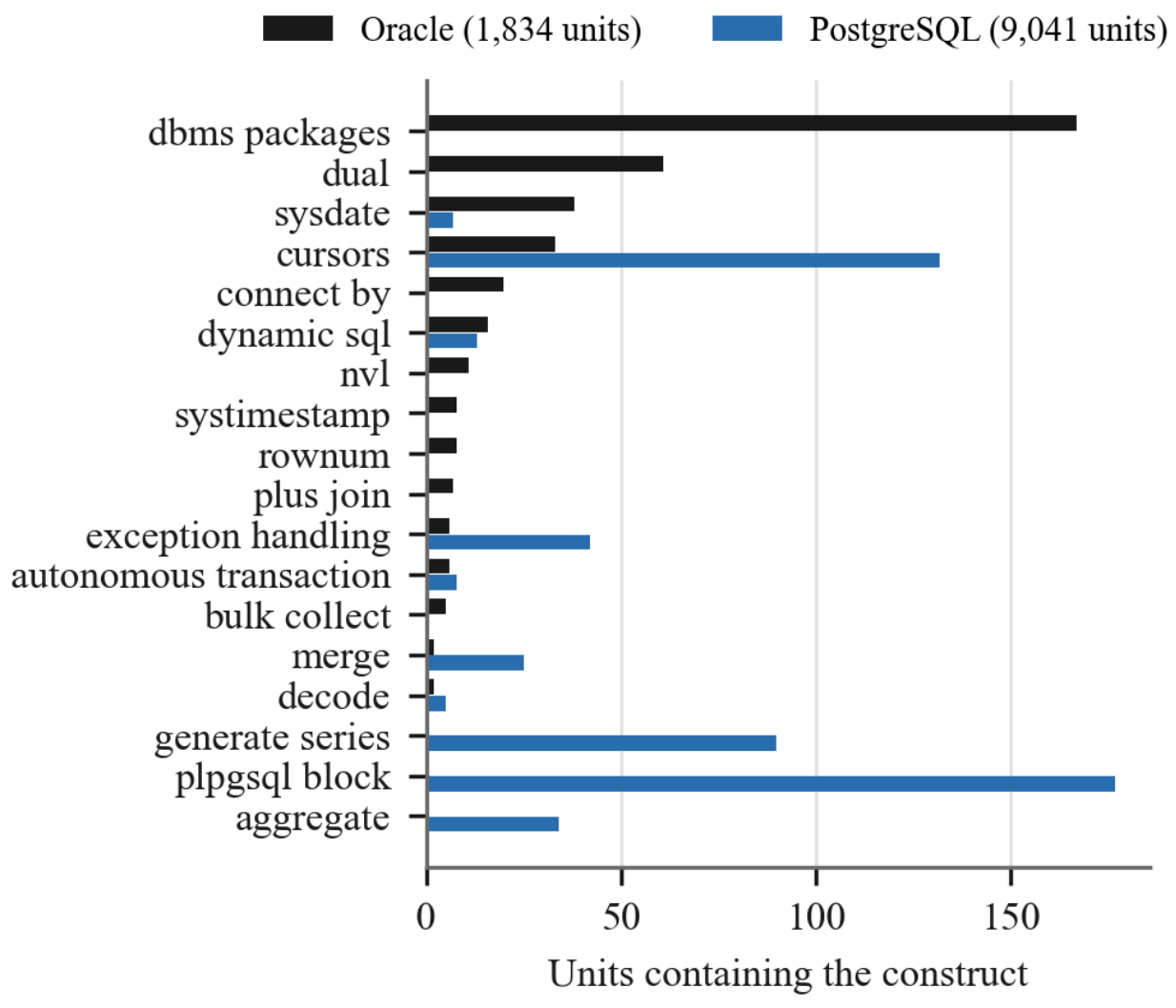

Fig. 2. Construct profile of the Study 1 corpus (units containing each detected construct).

The two sides therefore describe different features, not the same programs. Only DECODE, MERGE, dynamic SQL, autonomous transactions and SYSDATE occur on both sides in comparable numbers.

### C. Alignment and Evidence Tiers

Oracle units were matched to PostgreSQL units by a hybrid score of BM25, typed AST features and structural similarity. This produced 1,833 candidate pairs. Every pair was recorded as a 1:1 correspondence, but the pairs point to only 664 distinct PostgreSQL targets. Fig. 3 shows how often a target is reused: 352 targets serve one source each, while 28 targets serve more than ten sources and together cover 517 pairs.

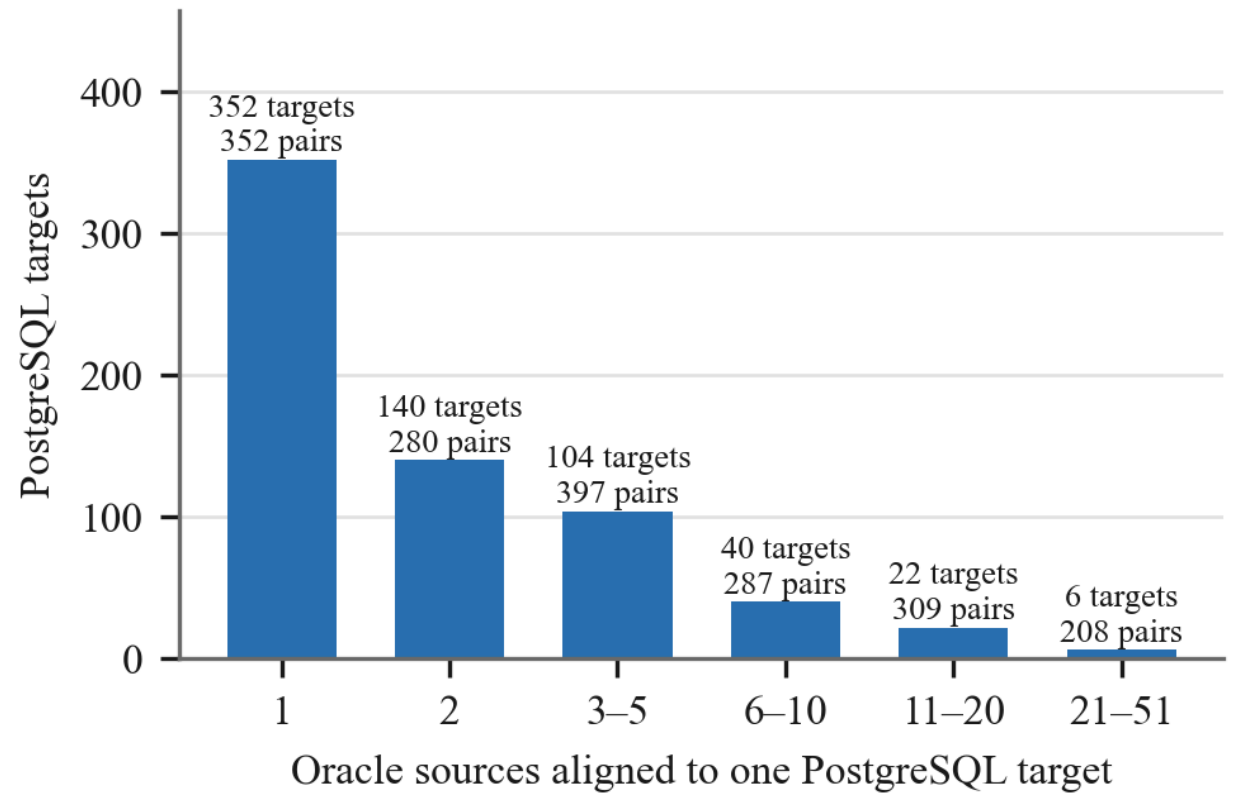

Fig. 3. Number of Oracle sources aligned to the same PostgreSQL target (1,833 pairs, 664 targets).

A typical pair links a short Oracle statement to a PostgreSQL documentation example on the same topic, not to its translation. Tiers were assigned automatically (Table IV). G1 means confirmed provenance and a parsable static reference: the documentation is human-written, but no pair was reviewed. All 1,677 pairs in the adjudication queue were still open at the time of the runs.

TABLE IV. Pair Evidence Tiers and Alignment Audit

| Tier or check | Evidence | Count |
|---|---|---|
| G0 | Aligned or generated; no correctness evidence | 504 |
| G1 | Confirmed provenance, parsable static reference | 1,329 |
| G2 | G1 plus PostgreSQL deployment checks | 0 |
| G3 | G2 plus behavioural check against Oracle | 0 |
| Distinct targets | PostgreSQL targets used by the 1,833 pairs | 664 |
| Reviewed pairs | Pairs checked by a person | 0 (queue 1,677) |

Because G2 and G3 are empty, every Study 1 claim is limited to structural agreement. The reuse of targets in Fig. 3 is also the path through which references later cross the train/test boundary.

### D. Migration Specification Construction

The proposed specification lifecycle defines two stages for constructing a structured, versioned migration specification in two separate stages (Fig. 4). Stage A, source-grounded extraction, does not see the target. It extracts typed interfaces, signatures, schema objects, input/output contracts inferable from code, SQL semantics, branches, side effects, reads and writes, transactional and exception behavior, dependencies, assumptions, Oracle-specific behavior, uncertainty, and non-inferable requirements. Missing business semantics are recorded as unknowns, not invented.

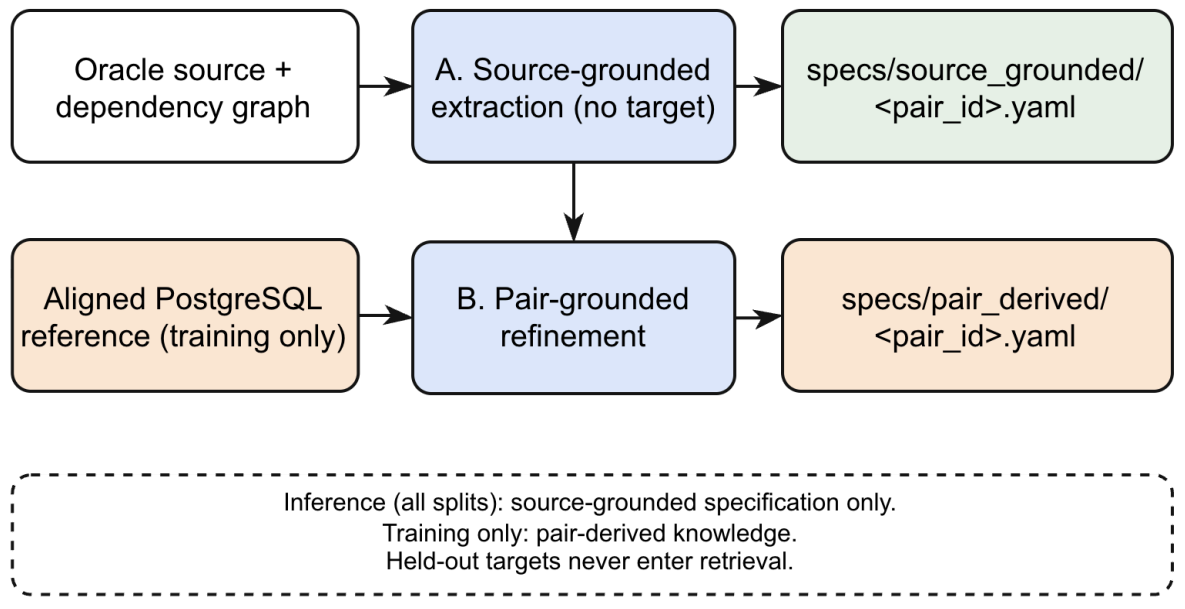

Fig. 4. Lifecycle of source-grounded and pair-derived specifications.

Stage B, pair-grounded refinement, applies to training pairs only. It compares the source-grounded specification with the aligned PostgreSQL target to record observed transformation decisions, alternatives, conflicts, confidence, and verification tier, and it never rewrites source requirements to fit the target. Two artifacts result: `specs/source_grounded/<pair_id>.yaml`, permissible at inference for all splits and derived only from source and graph, and `specs/pair_derived/<pair_id>.yaml`, training-only knowledge. The source-grounded schema contains the blocks `source`, `interfaces`, `semantics`, `dependencies`, `constraints`, `unknowns`, and `extraction`; the pair-derived schema contains `reference`, `observed_mappings`, `feedback`, and `review`.

Formally, the source-grounded specification $S_i^{src}$ is generated from Oracle code and dependency facts alone, whereas the training-only pair specification $S_i^{pair} = Augment(S_i^{src}, P_i^{ref})$ may summarize observed Oracle→PostgreSQL correspondences, supported idioms, and reference-specific annotations. A held-out candidate is generated from $S^{src}$ and admissible training-derived knowledge, never from its own target reference. Model-requested clarifications are tracked as unresolved assumptions, and a model-generated specification is not authoritative when it contradicts source facts. The generation prompt receives the source-grounded YAML, Oracle code, permitted dependency context, retrieved *training* evidence only, the output schema, and verification instructions. The held-out PostgreSQL target is never provided. Prompt, version, model, budget, and hash are stored in reproducible manifests, outputs must parse against a schema, and deterministic checks reject unsupported semantic claims. The specification is a *reconstruction* from code, which is explicitly distinguished from an authoritative business specification.

This specification lifecycle was not exercised in the runs reported here. The Stage A audit in the code base returned fixed placeholder scores, so we report no specification audit. The spec-first strategy used in B1 is a fixed checklist of Oracle constructs (NVL, VARCHAR2, NUMBER, SYSDATE, PL/SQL blocks, ROWNUM, DECODE and (+) joins), not a per-unit specification. The specification effect therefore cannot be separated in Study 1.

### E. Original Split

The configuration asked for a split by documentation family, but the split script did not apply it: every pair formed its own component, so the split was random at pair level. It produced 1,099 training, 367 validation and 367 test units; 109 of the 136 test families also occur in training. Retrieval excluded experiences of the same family at inference. Targets were not part of the grouping: 141 of the 236 distinct test targets also occur as training targets. Training used 311 of the 1,099 training units, in seven batches of up to 50. Section VIII-E describes the repaired split.

## V. Dataset-Oriented Agent Learning Method

### A. Experience Representation

An experience is a typed record with an identifier, the originating training pair, source-specification and dependency features, the selected strategy, a transformation rule, the observed outcome, the evidence tier, applicability conditions, contraindications, negative examples, a confidence value, the checkpoint at which it was created, and a knowledge version. Negative examples and contraindications are first-class fields, so that the memory distinguishes positive from negative transfer and does not assume that a G1 pair proves behavioral correctness.

In the implementation, an experience stores the agent's own generated PostgreSQL code for a training unit, not the reference. An experience counts as verified when its own verifier result is a binary success.

### B. Ten-Stage Learning Loop

Fig. 5 depicts the intended full loop; Study 1 implemented a reduced version without per-unit specifications or a code-level dependency graph (Section IV). For each task, the agent (1) observes source-derived features, the dependency neighborhood, and applicable training examples; (2) retrieves with hybrid lexical, AST, and graph signals plus optional semantic ranking, excluding the current held-out target, task aliases, and duplicates; (3) selects among direct, source-spec-first, example-assisted, and dependency-decomposed migration, logging scores and rationale; (4) generates candidate PostgreSQL without reference leakage; and (5) verifies with parser, schema, and static dependency checks, optionally with PostgreSQL deployment, comparing to allowed references for training pairs only.

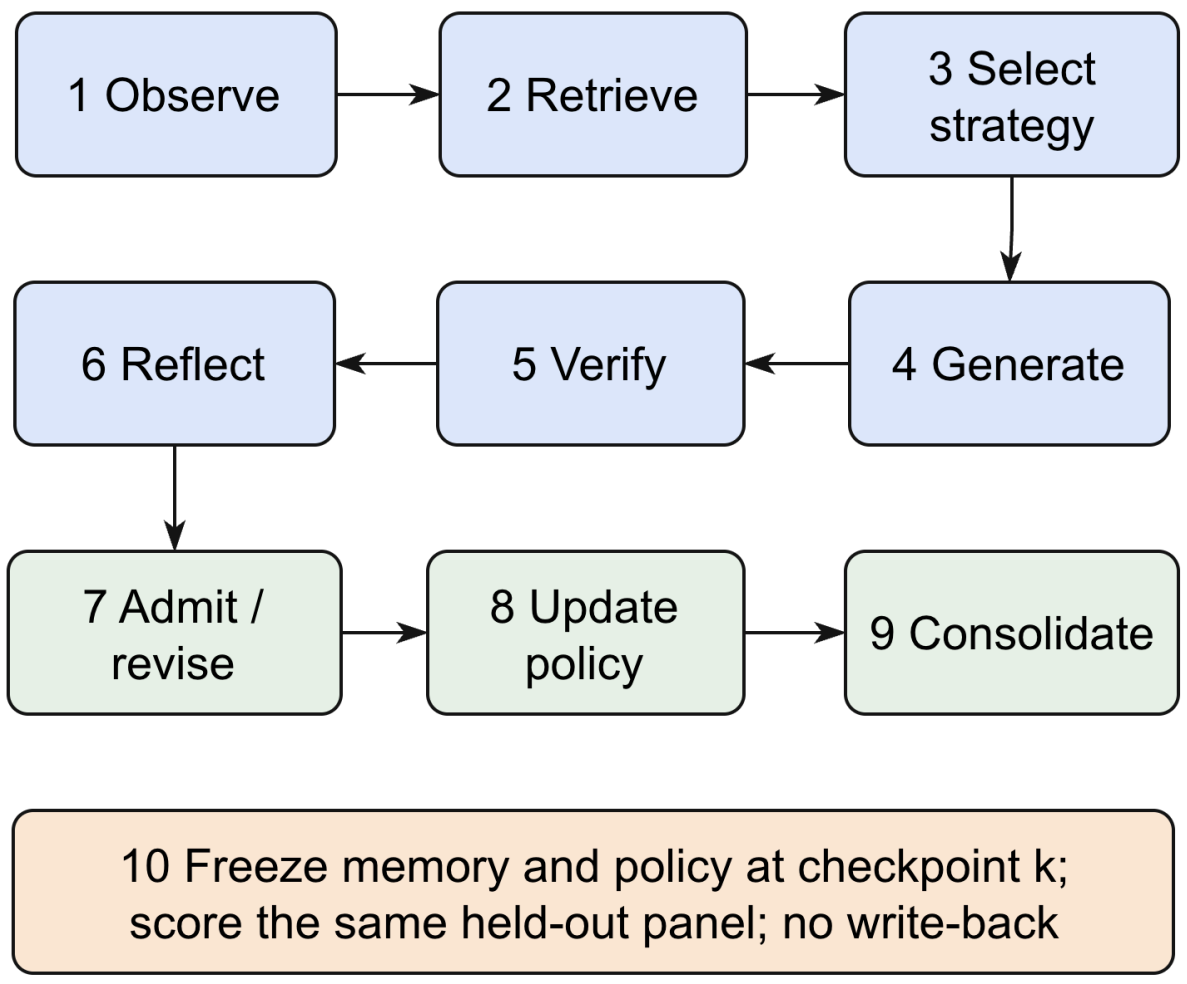

Fig. 5. Online experience and knowledge validation loop with checkpoint freezing.

The agent then (6) reflects by creating structured error and repair events grounded in verifier outputs rather than in unverified model self-evaluation; (7) admits or revises knowledge using evidence thresholds, applicability conditions, contradiction handling, feedback history, and deprecation; (8) updates the policy; (9) consolidates duplicate patterns while separating positive and negative transfer and storing versions and an audit trail; and (10) freezes memory and policy at a checkpoint and scores a stable held-out panel without writing its results back. Learning is made observable through `experience_events.jsonl`, `memory_snapshots/`, `policy_snapshots/`, `retrieval_traces.jsonl`, and `checkpoint_scores.csv`.

### C. Knowledge Admission and Consolidation

Admission is evidence-weighted. For a candidate rule $r$ with supporting and opposing verifier outcomes $s_j \in \{0, 1\}$ derived from experiences $e_j$, confidence is a smoothed, tier-weighted success rate:

$$c_r = \frac{\alpha + \sum_j \omega(e_j)\, s_j}{\alpha + \beta + \sum_j \omega(e_j)} \qquad (10)$$

where ω maps the evidence tier to a weight with $\omega(G0) \le \omega(G1) \le \omega(G2) \le \omega(G3)$, and $\alpha, \beta$ are prior pseudo-counts. A rule is admitted only if $c_r \ge \tau_c$, the support count reaches $\tau_n$, and it does not contradict an admitted rule of equal or higher tier without an explicit applicability split. Thresholds $\tau_c$, $\tau_n$, α, β, and the weights ω are fixed on validation data before any test run.

Extracted cases are normalized by source construct, target idiom, context features, and dependency signature,

and identical cases are deduplicated. Conflicting cases are not averaged; they retain evidence identifiers and validity scopes. Admission applies schema validation, source-grounding checks, reference-provenance checks, forbidden-pattern checks, and verification-tier gates. Cases supported only by G0–G1 evidence remain advisory and cannot justify claims of executable or behavioral correctness.

The thresholds were $\tau c = 0.65$ and $\tau n = 2$ with tier weights $\omega(G0) = 0.10$, $\omega(G1) = 0.50$, $\omega(G2) = 0.80$ and $\omega(G3) = 1.00$, and $\alpha = \beta = 1$. Verified experiences are retrieved as positive examples; an admitted rule adds a ranking bonus. Unverified experiences can be shown as counter-examples.

### D. Adaptive Strategy Policy

The policy is a simple, reproducible contextual strategy ranker over the four strategies. With features $\phi(x)$ of the source specification and dependency profile, strategy probabilities are

$$\pi_k(a|x) = \frac{exp\left(\theta_a^{\top}\phi(x)/\tau\right)}{\sum_{a'} exp\left(\theta_{a'}^{\top}\phi(x)/\tau\right)} \quad (11)$$

and parameters are updated on training batches by a baseline-subtracted policy-gradient step:

$$\theta_a \leftarrow \theta_a + \eta\,(R - b(x))\phi(x)\left(1\left[a_t = a\right] - \pi(a|x)\right) \quad (12)$$

with reward $R$ equal to static reference conformity minus a cost penalty, baseline $b(x)$, and learning rate $\eta$. A calibrated rule-priority model may replace this ranker if it performs better on validation; the choice is fixed before test runs. A fixed-policy memory-only variant (B5) uses the same memory and retrieval but no policy updates, which isolates adaptive policy learning from memory.

Training samples strategies with a seeded generator, as (12) assumes; held-out evaluation is greedy. The learning rate was $\eta = 0.05$, $\tau = 1$ and the cost weight 0.001. The untrained policy starts with a small bias towards the spec-first strategy, so P1 at checkpoint 0 behaves like B1.

### E. Algorithm Summary

Algorithm 1 summarizes the procedure. Inputs are the training and fixed test groups, source parser, reference analyzer, independent verifier, an initially empty memory $M_0$, and an initial policy $\pi_0$.

- (1) Audit and group-split all units.
- (2) Generate and validate source-only specifications.
- (3) For each training batch, generate pair specifications, extract cases, deduplicate, and evidence-label them.
- (4) For each training task, retrieve permitted cases, select a strategy, generate, verify, and log the outcome.
- (5) Admit evidence-qualified cases, revise the policy, and persist versioned snapshots.
- (6) At each checkpoint, freeze memory and policy and evaluate on held-out source-only tasks.
- (7) Compare paired outcomes with controls and report negative transfer.
- (8) Optionally run target deployment or independent behavior checks on an eligible subset.

Study 1 executed data splitting, retrieval, migration generation, static proxy checks, memory and policy updates, checkpoint evaluation, and the reported controls. However, the source-only and pair-derived specification lifecycle described in Steps (2)–(3) was not exercised: the Stage A audit returned placeholder scores (Section IV-D). No code-level dependency graph or independent Oracle execution evidence was available for Study 1. Study 3 separately examines execution-based verification.

## VI. Experimental Design

### A. Study 1: Systems Compared

All Study 1 systems use Qwen3-Coder-480B-A35B through Amazon Bedrock (qwen.qwen3-coder-480b-a35b-v1:0) at temperature 0, with the same instructions and one generation call per unit. Only B3 has a repair budget. Responses are cached on a hash of model, prompt and temperature; replays report the tokens of the first call. Table V lists the systems, ablations and controls. Ablations and controls reuse the final P1 memory and policy with one component removed at inference; none was retrained.

TABLE V. Baselines, Proposed System, Ablations and Controls

| ID | Configuration |
|---|---|
| B0 | Direct LLM translation; no checklist, no memory. |
| B1 | Spec-first construct checklist; no memory. |
| B2 | B1 + static retrieval of matched training pairs. |
| B3 | B1 + task-local repair loop; no cross-task learning. |
| B4 | B1 + unfiltered persistent memory; fixed policy. |
| B5 | B1 + verified persistent memory; fixed policy. |
| P1 | B5 + admission + adaptive strategy policy. |
| A1–A6 | P1 minus: admission; keyword-depth ranking term; memory; policy adaptation; case refinement; counter-examples. |
| C1–C4 | Controls: random memory, shuffled evidence, unrelated memory, equal-budget static retrieval. |
| M0–M3 | Model-free (Study 2): empty output; copy the Oracle source; copy the target of the most similar training source; copy a random training target. |

P1 versus B5 isolates policy learning. B5 versus B1 isolates memory. P1 versus B4 isolates admission. A4 applies the checkpoint-0 policy to the final memory and therefore equals B5 by construction.

### B. Chronological Protocol

Training units arrive in seven batches. After each batch, memory and policy are frozen and the same 367-unit panel is scored. Held-out outputs are never written back. Results are also split by tier (G0 and G1 only).

### C. Verification Ladder

Table VI lists the measurable proxies. In Study 1 only the static levels exist. Deployment and behavioural levels exist only in Study 3.

TABLE VI. MEASURABLE PROXIES UNDER THE AVAILABLE ASSETS

| Proxy | Definition and limit |
|---|---|
| PV | Static dialect check by regular expressions; a proxy, not a parser. |
| RC | Weighted share of four reference checks (Section III-B); agreement, not behaviour. |
| CC | Coverage of key SQL constructs of the reference; part of RC. |
| DS | Deployment success; Study 3 only. |
| BE | Behavioural agreement with Oracle output and errors; Study 3 only. |

The lower two rows matter for any claim about migration quality. Study 1 cannot reach them, which is why Study 3 is part of this article.

### D. Study 2: Leakage Audit, Model-Free Baselines and Repaired Split

The leakage audit and model-free comparisons in Study 2 use the logged outputs of Study 1 without new LLM calls; the final reference-grouped evaluation includes a separate LLM rerun (Section VIII-I). First, it marks each test unit whose PostgreSQL target also occurs as a training target. Second, it re-scores every system on the leak-free and the shared sub-panels, paired by group as in (13). Third, it scores model-free baselines with the same verifier: M0 returns nothing; M1 returns the Oracle source unchanged; M2 returns the PostgreSQL target of the training pair whose Oracle source is most similar by TF-IDF cosine; M3 returns a random training target. Fourth, a checkpoint monitor reads the memory snapshots and retrieval traces and reports, per checkpoint, the number of verified experiences, the distinct positive examples shown to the panel, the retrieval churn (mean Jaccard distance of retrieved sets between consecutive checkpoints) and the strategy mix. The monitor follows the idea of comparing what changed in the learned state with what changed in behaviour [36].

Fifth, Study 2 builds a split grouped by PostgreSQL target, tests it for shared targets and shared target text, re-runs the model-free baselines on it, re-scores the systems that use no training data (B0, B1, B3) on the new test units they already cover, estimates the power of a rerun, and reruns B1, B5 and P1 on the new split.

### E. Study 3: Execution-Judged Evaluation

Study 3 comes from the lab's migration agent, which translates Oracle PL/SQL with a production routing (DeepSeek-V4-Flash, escalating to DeepSeek-V4-Pro). These names identify the reported routing choices; the specific provider-resolved model release for each run is not documented in this manuscript. Its corpus has 34 Oracle files with 216 units, later 224. Each unit is installed in a PostgreSQL validation database in dependency order and run against scenarios. Oracle ground truth comes from captured Oracle transcripts and, for disputed cases, from execution on an Oracle Database Free container. A unit is demonstrably correct when it is accepted and Oracle's own output and errors do not contradict it, or when it fails with exactly the error Oracle raises. Every change to the prompt or harness went through a pre-registered fix cycle: one change per arm, predicates written before the run, and promote, revert or hold decided against the real artifacts. We also ran a do-nothing agent beside every reported number.

### F. Study 4: Knowledge Write-Back at Scale

Study 4 tests whether a single correction written back into a FAISS knowledge base becomes retrievable and stays retrievable. The gap case is an Oracle XMLTABLE query with no matching rule. The base holds 12 real migration rules, padded with synthetic filler rules to 120 and 1,200 entries. We record top-3 retrieval before and after ingesting the correction and the distance margin between the correct rule and the best irrelevant one, first with all-MiniLM-L6-v2 and then with the production embedding model BGE-M3-SQL. A second test ingests seven corrections one by one and re-checks all seven gap cases after each step.

### G. Hypotheses and Statistical Plan

Table VII lists the hypotheses and their outcomes. The unit of analysis is the independent group. Intervals come from 10,000 group-bootstrap resamples with seed 1; binary contrasts use the exact McNemar test with Holm correction. The configuration fixes a smallest effect of interest (SEI) of 2.0 percentage points. We cannot document that it was frozen before the runs, so verdicts are interval-based; we also state whether the threshold was met. Group-level paired contrasts are not necessarily equal to the differences between the unit-weighted mean RC values shown in the summary tables.

TABLE VII. HYPOTHESES AND OUTCOMES

| ID | Hypothesis | Outcome |
|---|---|---|
| H1 | P1 beats B5 on held-out RC (policy learning beyond memory). | Not supported: −0.004 [−0.010, 0.001]; leak-free −0.003 [−0.010, 0.004]; new split 0.000. |
| H2 | P1 has less negative transfer than B4. | Not testable: B1 passes 1 of 367 units, so NTR is 0 or 1/367. |
| H3 | Graph-aware learning helps dependency-heavy tasks. | Not testable: no dependency graph; keyword proxy −0.003 [−0.008, 0.000]. |
| H4 | P1 improves validated outcomes per token. | Not supported: 83,382 vs 40,687 tokens per pass. |
| H5 | Learned patterns transfer to unseen construct families. | Inconclusive: P1 − B1 = 0.006 [−0.008, 0.020], 18 units. |

The paired effect for H1 is

$$\Delta = \frac{1}{G}\sum_{g=1}^{G}\left(q_g^{P1} - q_g^{B5}\right) \quad (13)$$

with a confidence interval from bootstrap resampling over groups. No effect in Study 1 reached the 2.0-point threshold.

## VII. STUDY 1: AGENT LEARNING ON THE DOCUMENTATION CORPUS

### A. Baseline Comparison

Table VIII compares the systems on the frozen panel at the final checkpoint. Every memory system scored above the no-memory baseline B1 (0.143). Verified memory B5 reached 0.161 and unfiltered memory B4 0.163, so admission made no difference. The adaptive agent P1 reached 0.155, below B5. Binary success was rare: at most 8 of 367 units passed. Task-local repair (B3) raised the dialect proxy from 0.749 to 0.807 but not RC.

TABLE VIII. Results on the Frozen Test Panel (367 Units, 136 Groups, Final Checkpoint)

| System | RC (95% CI) | Dialect proxy | Passes | NTR vs B1 |
|---|---|---|---|---|
| B0 | 0.133 [0.119, 0.147] | 0.597 | 0 | 0.003 |
| B1 | 0.143 [0.129, 0.155] | 0.749 | 1 | reference |
| B2 | 0.152 [0.136, 0.168] | 0.728 | 7 | 0.000 |
| B3 | 0.139 [0.127, 0.151] | 0.807 | 1 | 0.003 |
| B4 | 0.163 [0.148, 0.178] | 0.738 | 6 | 0.003 |
| B5 | 0.161 [0.146, 0.176] | 0.711 | 8 | 0.000 |
| P1 | 0.155 [0.141, 0.169] | 0.684 | 3 | 0.003 |

Fig. 6 splits the same results by tier. G1 units score higher than G0 units for every system, for example 0.173 against 0.123 for B5, and the order of the systems is the same in both tiers.

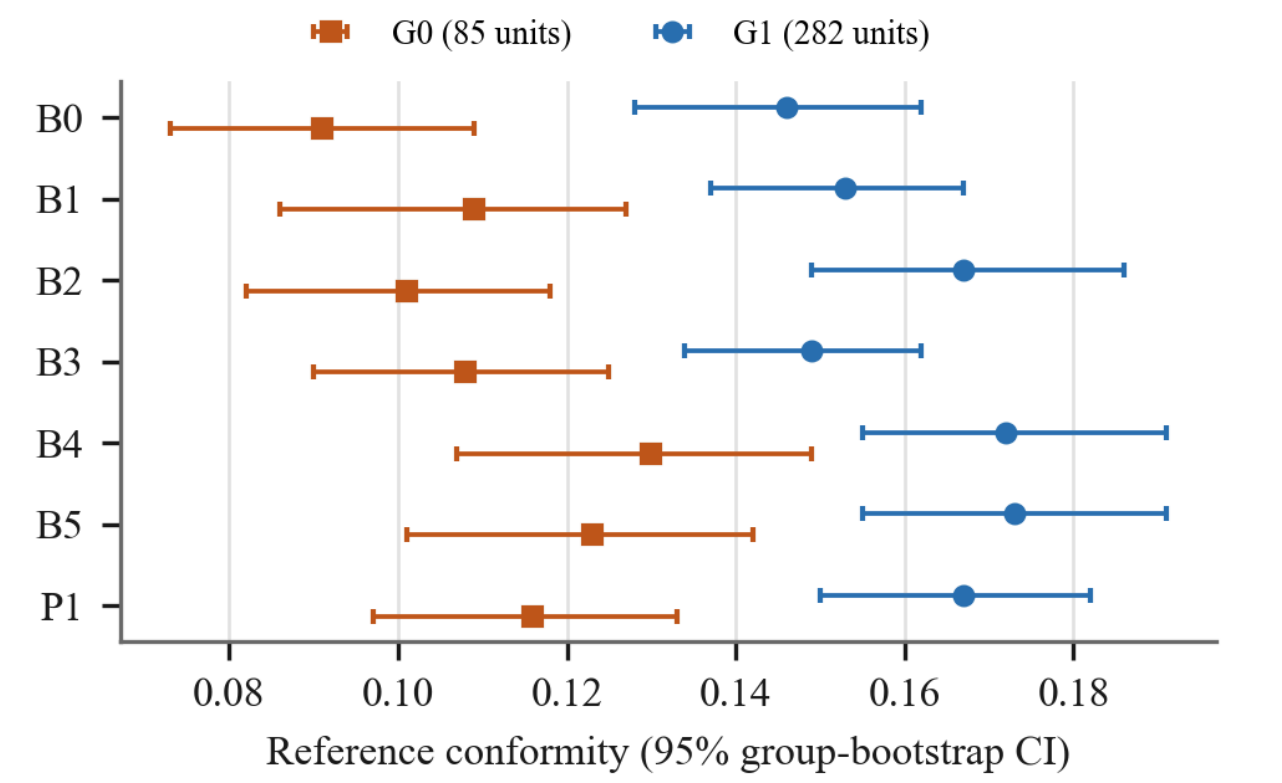

Fig. 6. Reference conformity by evidence tier (G0: 85 units in 53 groups; G1: 282 units in 118 groups).

The tier gap is larger than any gap between systems. Better references raise the score more than any part of the agent does, which already hints that RC rewards the reference as much as the output.

## B. Continual-Learning Checkpoints

Table IX and Fig. 7 follow P1 across checkpoints. RC rose from 0.141 to 0.155, and the paired gain over checkpoint 0 was 0.012 [0.004, 0.022] at k = 7. NTR stayed at 1/367 until k = 6 and was 2/367 at k = 7.

TABLE IX. Checkpoints of P1 on the Frozen Panel, with Monitor Readings

| k | Mem. | Verif. | RC | LG(k) (95% CI) | Strategy | Churn |
|---|---|---|---|---|---|---|
| 0 | 0 | 0 | 0.141 | — | spec-first | — |
| 1 | 50 | 0 | 0.145 | 0.005 [-0.003, 0.013] | spec-first | 1.00 |
| 2 | 100 | 0 | 0.151 | 0.008 [-0.001, 0.019] | spec-first | 0.63 |
| 3 | 150 | 0 | 0.150 | 0.007 [-0.002, 0.018] | spec-first | 0.10 |
| 4 | 200 | 0 | 0.150 | 0.007 [-0.002, 0.018] | spec-first | 0.02 |
| 5 | 250 | 0 | 0.152 | 0.013 [0.003, 0.024] | example | 0.55 |
| 6 | 300 | 0 | 0.153 | 0.013 [0.003, 0.025] | example | 0.02 |
| 7 | 311 | 1 | 0.155 | 0.012 [0.004, 0.022] | example | 0.51 |

The curve in Fig. 7 looks like learning: it rises after the first batches and again after k = 5.

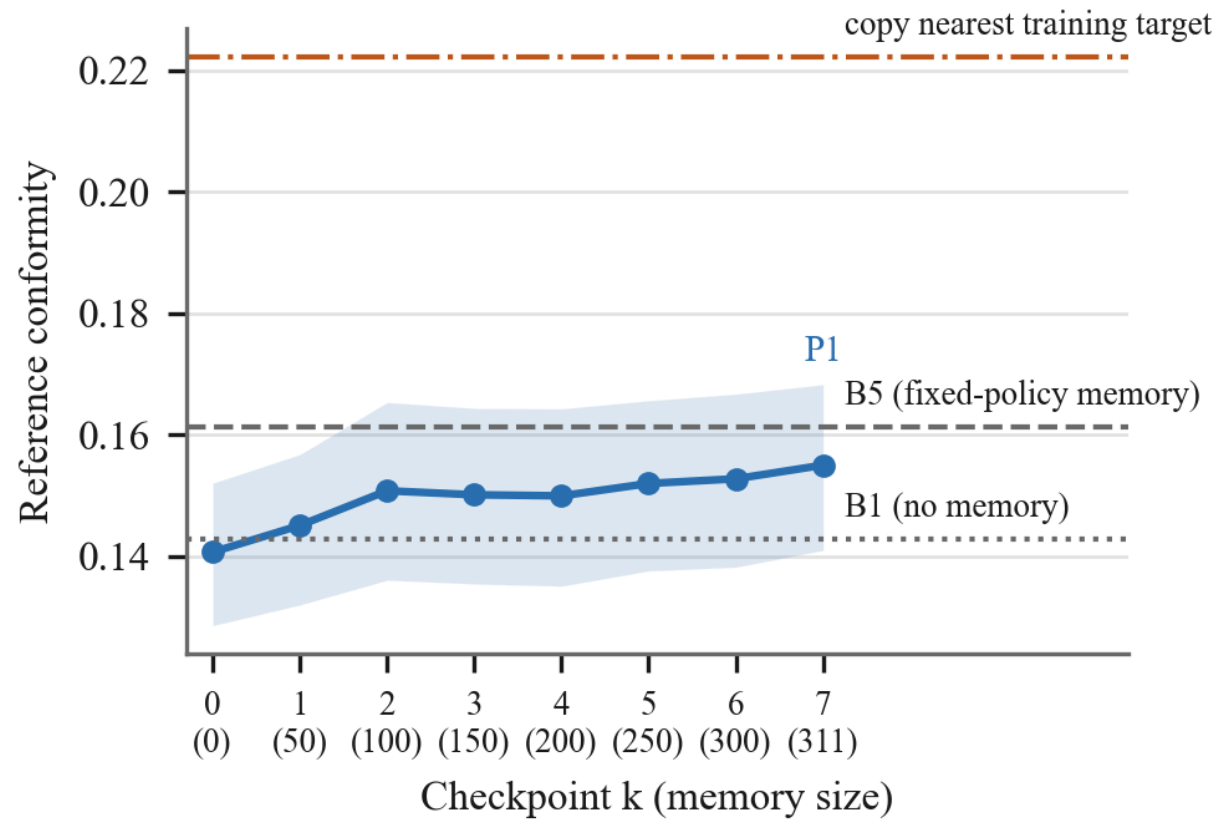

Fig. 7. Reference conformity of P1 by checkpoint with a 95% group-bootstrap band, against B1, B5 and the model-free copy baseline M2.

The checkpoint monitor in Fig. 8 shows what lies behind this curve. Until k = 6 the memory held no verified experience, so the panel saw only counter-examples. At k = 5 the greedy policy switched every unit at once from spec-first to example-assisted; strategy entropy stayed at zero throughout, so the policy never used context. At k = 7 a single verified experience appeared, and 356 of 367 units (97%) received it as their only positive example.

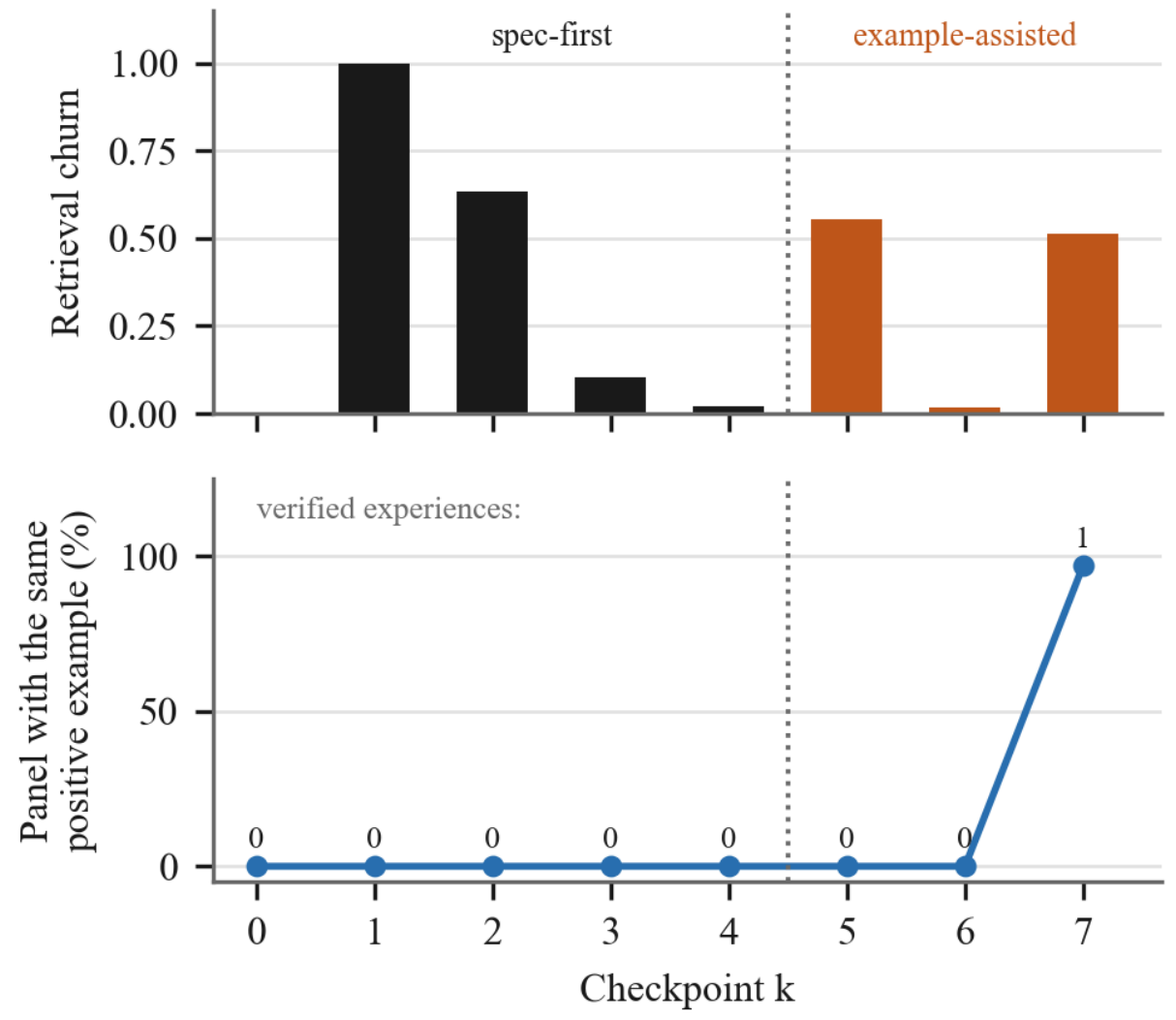

Fig. 8. Checkpoint monitor of P1: retrieval churn coloured by the strategy used on the whole panel (top), and the share of the panel that receives the same positive example, with the number of verified experiences (bottom).

No rule was ever admitted. During the audit we also found that the example-assisted prompt rendered counter-examples without their counter-example label; the code is now fixed for the rerun. The rise at k = 5 and k = 7 therefore coincides with a global strategy switch and with one shared example, not with growing knowledge.

## C. Subgroups

Table X splits the final gain. The dependency-depth field in the logs is a keyword count (EXECUTE, CALL, SELECT … FROM, INSERT INTO and similar), so the split is a proxy for statement content, not for dependencies.

TABLE X. FINAL-CHECKPOINT GAIN BY SUBGROUP

| Subgroup | Groups/units | LG (95% CI) | RU proxy (95% CI) |
|---|---|---|---|
| Keyword count ≥ 1 | 63/131 | -0.005 [-0.013, 0.001] | -0.001 [-0.006, 0.003] |
| Keyword count 0 | 98/236 | 0.023 [0.011, 0.036] | 0.028 [0.017, 0.041] |
| Rare constructs | 21/28 | 0.008 [0.001, 0.017] | 0.006 [0.001, 0.013] |
| Unseen families | 14/18 | 0.011 [0.000, 0.022] | 0.008 [0.000, 0.018] |
| Tier G0 | 53/85 | 0.003 [-0.006, 0.013] | 0.013 [0.003, 0.023] |
| Tier G1 | 118/282 | 0.016 [0.005, 0.029] | 0.019 [0.010, 0.029] |

The gain sat in units with a zero count, 0.023 [0.011, 0.036], and in G1 units, 0.016 [0.005, 0.029]. Units with a non-zero count gained nothing. RU is a memory-level proxy, P1 minus A3 on units with a non-empty retrieval trace; per-rule counterfactuals (9) were not run.

## D. Cost

Table XI reports tokens, calls and repairs per unit from the run manifest. Memory made prompts longer (B5: 887 tokens per unit against 521 for B1).

TABLE XI. TOKENS, CALLS AND REPAIRS PER UNIT

| System | Tokens | Calls | Repairs | Passes | Tokens/pass |
|---|---|---|---|---|---|
| B0 | 206.6 | 1.00 | 0.00 | 0 | n/a |
| B1 | 521.3 | 1.00 | 0.00 | 1 | 191,304 |
| B2 | 782.0 | 1.00 | 0.00 | 7 | 40,999 |
| B3 | 792.5 | 1.67 | 0.67 | 1 | 290,831 |
| B4 | 880.4 | 1.00 | 0.00 | 6 | 53,848 |
| B5 | 886.9 | 1.00 | 0.00 | 8 | 40,687 |
| P1 | 681.6 | 1.00 | 0.00 | 3 | 83,382 |

P1 used fewer tokens than B5 but also passed fewer units, so its cost per pass was twice that of B5. Mean latency per unit was 3.35 s for B5 and 3.28 s for P1. Costs of retrieval, consolidation and validation were not instrumented, and the per-token price is not known, so no currency cost is given.

## E. Ablations and Controls

Table XII removes one component at a time from the final P1 state.

TABLE XII. ABLATIONS AND CONTROLS ON THE FINAL P1 STATE

| Configuration | RC | ΔRC vs. P1 (95% CI) |
|---|---|---|
| A1 – admission/validator | 0.156 | 0.001 [-0.004, 0.007] |
| A2 – graph context | 0.157 | 0.001 [0.000, 0.002] |
| A3 – persistent memory | 0.136 | -0.017 [-0.027, -0.010] |
| A4 – policy adaptation | 0.161 | 0.004 [-0.001, 0.010] |
| A5 – case refinement | 0.151 | -0.002 [-0.009, 0.005] |
| A6 – failure examples | 0.156 | 0.006 [-0.004, 0.017] |
| C1 random memory | 0.142 | -0.014 [-0.027, -0.003] |
| C2 shuffled evidence | 0.140 | -0.010 [-0.022, -0.000] |
| C3 unrelated memory | 0.145 | -0.011 [-0.020, -0.004] |
| C4 equal-token-budget retrieval | 0.146 | -0.008 [-0.023, 0.006] |

Only removing memory (A3) hurt RC, by 0.017 [0.010, 0.027] relative to P1. Removing admission (A1), the keyword-depth term (A2), case refinement (A5) or counter-examples (A6) changed nothing beyond noise. Random, shuffled and unrelated memory (C1–C3) scored 0.010 to 0.014 below P1. So the content of memory mattered, but none of the mechanisms that select or grade it did.

## F. Failure Taxonomy and Case Traces

Table XIII classifies failing units automatically from verifier outputs; no manual labelling was done.

TABLE XIII. FAILURE TAXONOMY (AUTOMATIC CATEGORIES)

| Category | B5 | P1 |
|---|---|---|
| Dialect-proxy failure | 106 | 115 |
| Low conformity (RC < 0.5) with dialect pass | 253 | 248 |
| Negative transfer from memory | 0 | 1 |
| Unresolved dependency | not logged | not logged |

Most failures were low-conformity outputs that passed the dialect check. One P1 failure counts as negative transfer: B1 succeeded on it, and P1 failed with a retrieved example. The case traces explain the largest paired gains. The three units with the largest gain all have the target psql_docs_full_122, and all received the same verified experience, whose training unit has that same target. Their RC rose from 0.07–0.16 to 0.64–0.72.

# VIII. STUDY 2: LEAKAGE AUDIT AND REPAIRED SPLIT

## A. Shared References

Of the 236 distinct references in the test panel, 141 also occur as training targets. They cover 264 of the 367 test units (72%) in 105 groups; the other 103 units in 65 groups are leak-free. Memory stores the agent's own outputs, so a reference reaches the panel when a training output reproduces it closely. That is what happened with the single verified experience in Section VII-B.

## B. Re-Scoring on Leak-Free and Shared Units

Table XIV gives RC on the full, leak-free and shared panels.

TABLE XIV. Reference Conformity on the Full, Leak-Free and Shared Panels

| System | Full (367) | Leak-free (103) | Shared (264) |
|---|---|---|---|
| B0 | 0.133 [0.119, 0.147] | 0.140 [0.119, 0.158] | 0.131 [0.113, 0.148] |
| B1 | 0.143 [0.129, 0.155] | 0.146 [0.123, 0.165] | 0.142 [0.126, 0.156] |
| B2 | 0.152 [0.136, 0.167] | 0.136 [0.115, 0.155] | 0.158 [0.138, 0.178] |
| B3 | 0.139 [0.127, 0.151] | 0.150 [0.129, 0.167] | 0.135 [0.119, 0.150] |
| B4 | 0.163 [0.148, 0.178] | 0.152 [0.131, 0.170] | 0.167 [0.148, 0.187] |
| B5 | 0.161 [0.147, 0.176] | 0.148 [0.126, 0.167] | 0.166 [0.148, 0.186] |
| P1 | 0.155 [0.141, 0.168] | 0.147 [0.126, 0.166] | 0.158 [0.140, 0.176] |
| A3 | 0.136 [0.122, 0.148] | 0.133 [0.110, 0.154] | 0.137 [0.121, 0.153] |
| C1 | 0.142 [0.128, 0.155] | 0.141 [0.120, 0.160] | 0.142 [0.126, 0.159] |
| C3 | 0.145 [0.131, 0.157] | 0.150 [0.129, 0.167] | 0.143 [0.127, 0.158] |
| M0 empty | 0.000 [0.000, 0.000] | 0.000 [0.000, 0.000] | 0.000 [0.000, 0.000] |
| M1 copy source | 0.111 [0.096, 0.124] | 0.123 [0.100, 0.144] | 0.106 [0.091, 0.121] |
| M2 copy nearest target | 0.222 [0.185, 0.262] | 0.109 [0.088, 0.129] | 0.267 [0.218, 0.316] |
| M2 without shared targets | 0.104 [0.092, 0.117] | 0.101 [0.083, 0.121] | 0.105 [0.092, 0.119] |

Memory systems score higher on shared units than on leak-free ones, while systems without memory do not. Table XV and Fig. 9 give the paired contrasts behind this pattern.

TABLE XV. Paired Contrasts by Panel (Eq. 13, 95% Group-Bootstrap CI)

| Contrast | Full | Leak-free | Shared |
|---|---|---|---|
| B5 − B1 | 0.015 [0.004, 0.025] | 0.002 [-0.008, 0.011] | 0.023 [0.008, 0.039] |
| B4 − B1 | 0.015 [0.005, 0.027] | 0.002 [-0.008, 0.011] | 0.022 [0.008, 0.038] |
| B2 − B1 | 0.012 [-0.001, 0.027] | -0.005 [-0.017, 0.007] | 0.020 [0.004, 0.038] |
| P1 − B5 | -0.004 [-0.010, 0.001] | -0.003 [-0.010, 0.004] | -0.007 [-0.015, 0.001] |
| P1 − B1 | 0.010 [0.000, 0.020] | -0.001 [-0.010, 0.008] | 0.016 [0.004, 0.030] |
| P1 − M2 | -0.044 [-0.082, -0.007] | 0.028 [0.006, 0.050] | -0.075 [-0.126, -0.029] |

On shared units, verified memory beat no memory by 0.023 [0.008, 0.039]. On leak-free units the same contrast was 0.002 [−0.008, 0.011], and static retrieval was 0.005 below B1.

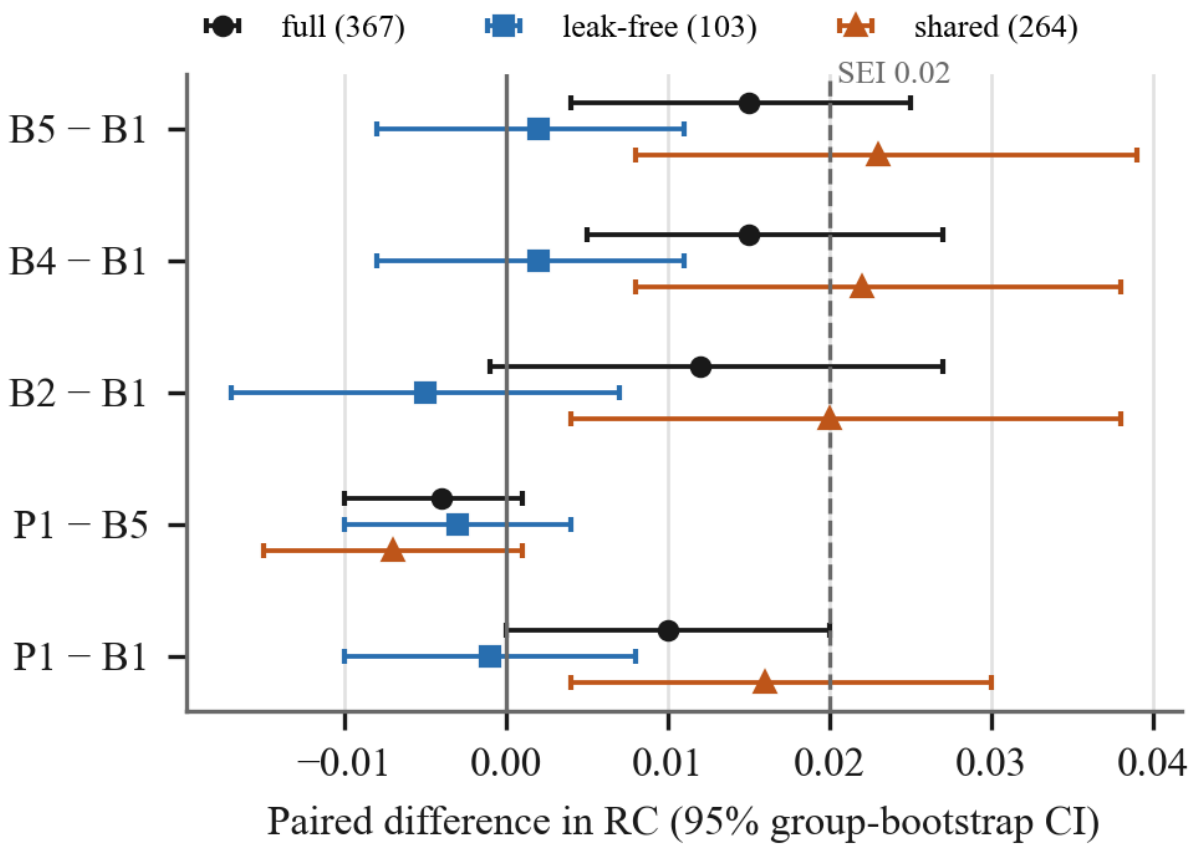

Fig. 9. Paired contrasts on the full, leak-free and shared panels; the dashed line marks the smallest effect of interest.

In Fig. 9 every memory contrast on the leak-free panel crosses zero, and only shared units reach the 2.0-point threshold. The adaptive policy did not help on any panel.

## *C. Model-Free Baselines*

A program that copies the PostgreSQL target of the most similar training source (M2) reached RC 0.222 [0.185, 0.262], above every LLM system. It passed 38 of 367 units (10.4%) against at most 8 for any system, and for 47 units (12.8%) it returned the exact reference. Fig. 10 puts the copy baselines next to the agent on both panels.

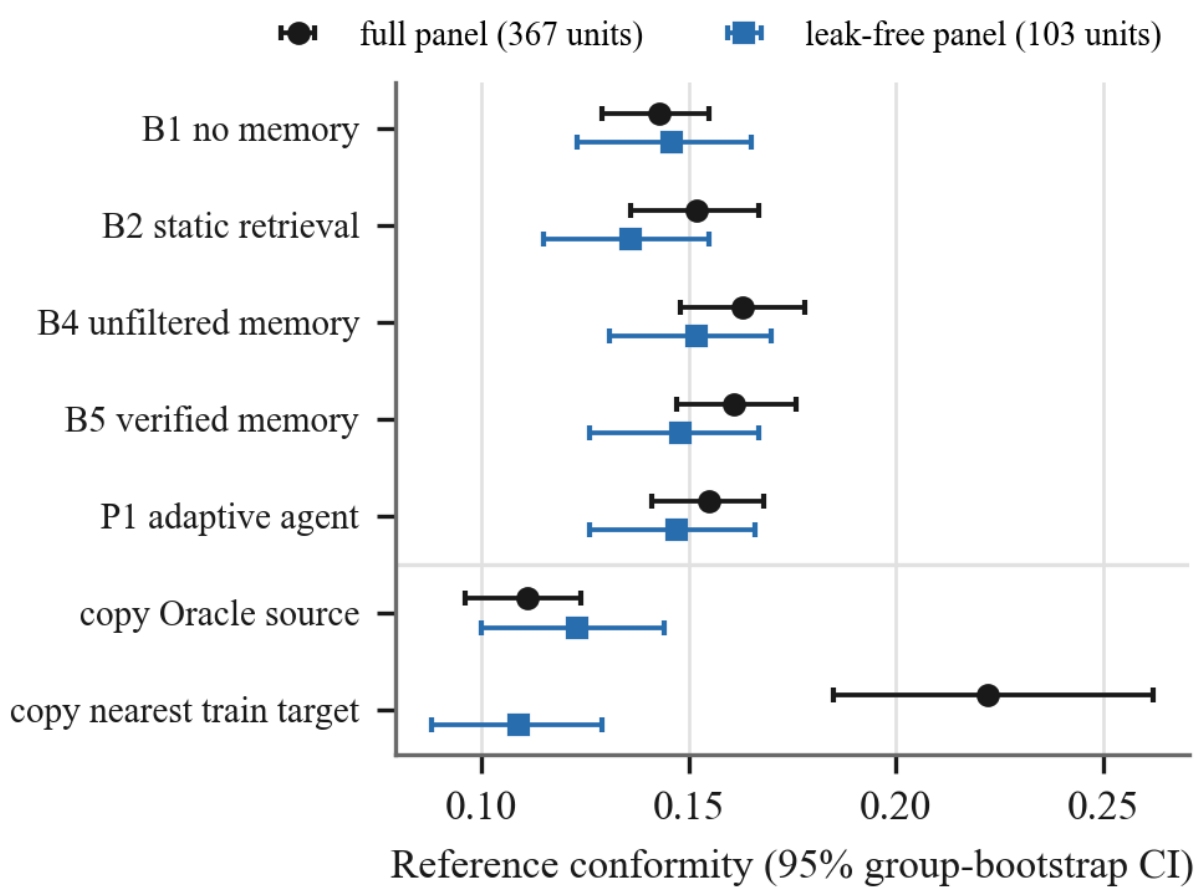

Fig. 10. Reference conformity on the full and leak-free panels for selected systems and the model-free copy baselines.

On leak-free units M2 fell to 0.109, below every LLM system, and P1 beat it by 0.028 [0.006, 0.050]. With shared targets removed from the pool, M2 scored 0.104 on all units. Returning the Oracle source unchanged (M1) scored 0.111, three quarters of B1, because RC rewards shared tokens and keywords. Returning nothing (M0) scores zero, so RC does not reward silence.

## *D. What the Audit Changes*

Without the audit, Study 1 would read as a small positive memory effect with a learning curve. With it, the memory effect is explained by reference reuse, the curve by one verified experience and a global strategy switch, and the best number on the panel by copying. None of the original leakage tests failed: they checked sources and pair identifiers, and the leak went through targets.

## E. A Split Grouped by Reference

We rebuilt the split with the PostgreSQL target as the unit of grouping. Grouping by target and documentation family together is not feasible: it joins 1,754 of the 1,833 pairs (96%) into one component. Grouping by target alone gives 664 components, the largest with 51 pairs (2.8%). Components were shuffled with seed 1 and assigned to the split with the largest deficit against 60/20/20. Table XVI compares the two splits.

TABLE XVI. ORIGINAL AND REFERENCE-GROUPED SPLITS

| Property | Original | Reference-grouped |
|---|---|---|
| Train / validation / test units | 1,099 / 367 / 367 | 1,097 / 362 / 374 |
| Distinct test targets | 236 | 118 |
| Test targets also in training | 141 | 0 |
| Test units with a shared target | 264 (72%) | 0 |
| Test families also in training | 109 of 136 | 88 of 113 |
| Test tiers G0 / G1 | 85 / 282 | 101 / 273 |
| Target-leakage tests passed | 0 of 6 | 6 of 6 |

A new test, tests/test_target_leakage.py, checks every pair of splits for shared target identifiers and for identical target text. It fails all six checks on the original split and passes them on the new one, together with the two original split tests. Documentation families still cross the split, as they did before; target grouping removes the reference path but not topical similarity.

## F. Model-Free Baselines on the New Split

Table XVII repeats the model-free baselines on both splits, with targets as bootstrap groups. Fig. 11 shows the same numbers.

TABLE XVII. MODEL-FREE BASELINES ON THE ORIGINAL AND REFERENCE-GROUPED SPLITS

| Baseline | Split | RC (95% CI) | Passes | Exact |
|---|---|---|---|---|
| M0 empty | original | 0.000 [0.000, 0.000] | 0 | 0 |
| M1 copy source | original | 0.111 [0.097, 0.124] | 0 | 0 |
| M2 copy nearest target | original | 0.222 [0.182, 0.264] | 38 | 47 |
| M3 random training target | original | 0.087 [0.074, 0.101] | 3 | 3 |
| M0 empty | grouped | 0.000 [0.000, 0.000] | 0 | 0 |
| M1 copy source | grouped | 0.108 [0.085, 0.133] | 0 | 0 |
| M2 copy nearest target | grouped | 0.119 [0.104, 0.137] | 4 | 0 |
| M3 random training target | grouped | 0.078 [0.068, 0.086] | 1 | 0 |

Copying the nearest training target fell from 0.222 to 0.119, and its exact copies of the reference fell from 47 to 0. Copying the source (M1) and a random target (M3) barely changed.

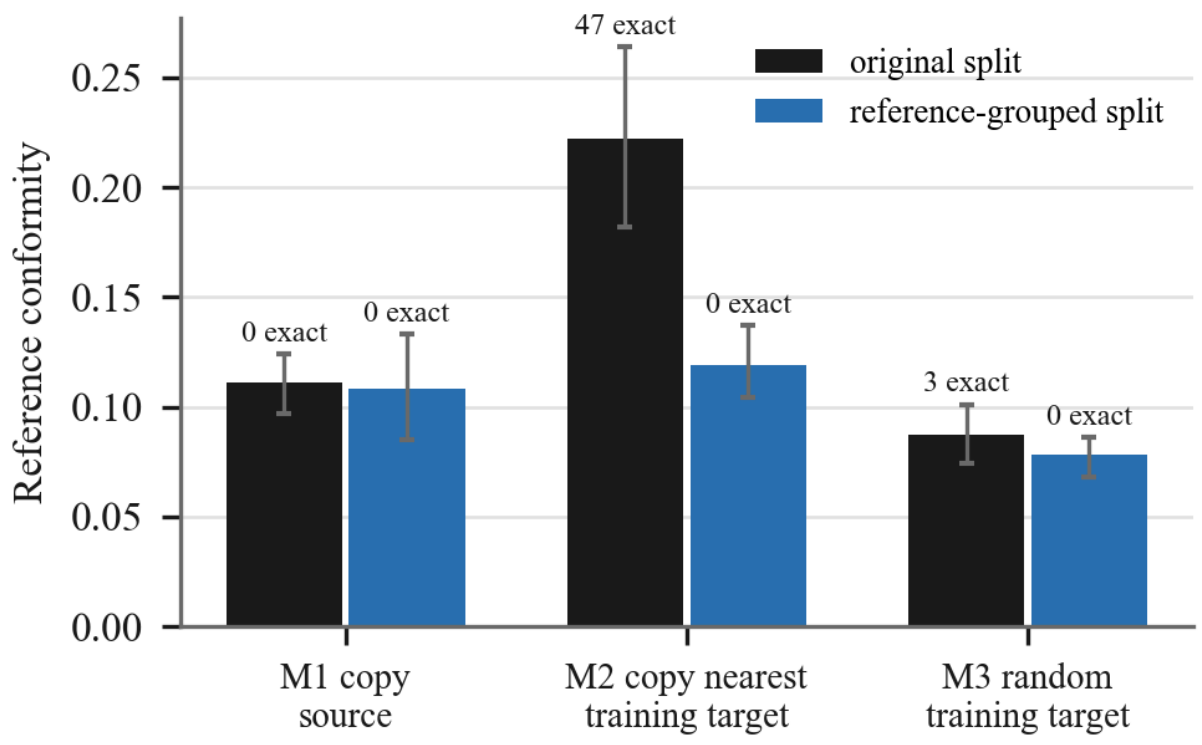

Fig. 11. Model-free baselines on the original and the reference-grouped split (labels give exact copies of the reference).

The remaining 0.119 is the floor that any LLM system on the new split must clear: it is what topical similarity alone earns under RC. It sits above M3 (0.078), so nearest-source retrieval still carries some signal without the reference path.

## G. Training-Independent Systems on the New Panel

B0, B1 and B3 use no training data, so their logged outputs stay valid under any split. Of the 374 new test units, 82 (49 targets) were also in the original panel and have outputs for these systems (Table XVIII).

TABLE XVIII. TRAINING-INDEPENDENT SYSTEMS ON THE 82 LOGGED UNITS OF THE NEW TEST PANEL

| System | Units | Targets | RC (95% CI) | Passes |
|---|---|---|---|---|
| B0 | 82 | 49 | 0.165 [0.125, 0.200] | 0 |
| B1 | 82 | 49 | 0.156 [0.121, 0.185] | 0 |
| B3 | 82 | 49 | 0.148 [0.121, 0.177] | 1 |
| M2 copy nearest target | 82 | 49 | 0.110 | – |
| M1 copy source | 82 | 49 | 0.107 | – |

On these units the no-memory baseline B1 scored 0.156 and copying scored 0.110. Without shared references, an LLM with a fixed checklist beats copying by about 0.05. Any memory gain in the rerun has to come on top of B1, not on top of copying.

## H. Power of the Rerun

The memory systems B2, B4, B5 and P1 depend on training data and must be rerun. Table XIX estimates the minimum detectable effect (MDE) at a two-sided α of 0.05 and 80% power from the SD of group-level paired differences in the logged run, grouped by target. Fig. 12 plots the same estimate against the number of groups.

TABLE XIX. Minimum Detectable Effect for the Rerun (α = 0.05, Power 0.80)

| Contrast | SD (groups) | MDE, 118 groups | MDE, 236 groups | Groups for 0.02 |
|---|---|---|---|---|
| B5 − B1 | 0.069 | 0.0178 | 0.0126 | 94 |
| P1 − B5 | 0.045 | 0.0115 | 0.0082 | 40 |
| P1 − B1 | 0.063 | 0.0162 | 0.0115 | 78 |

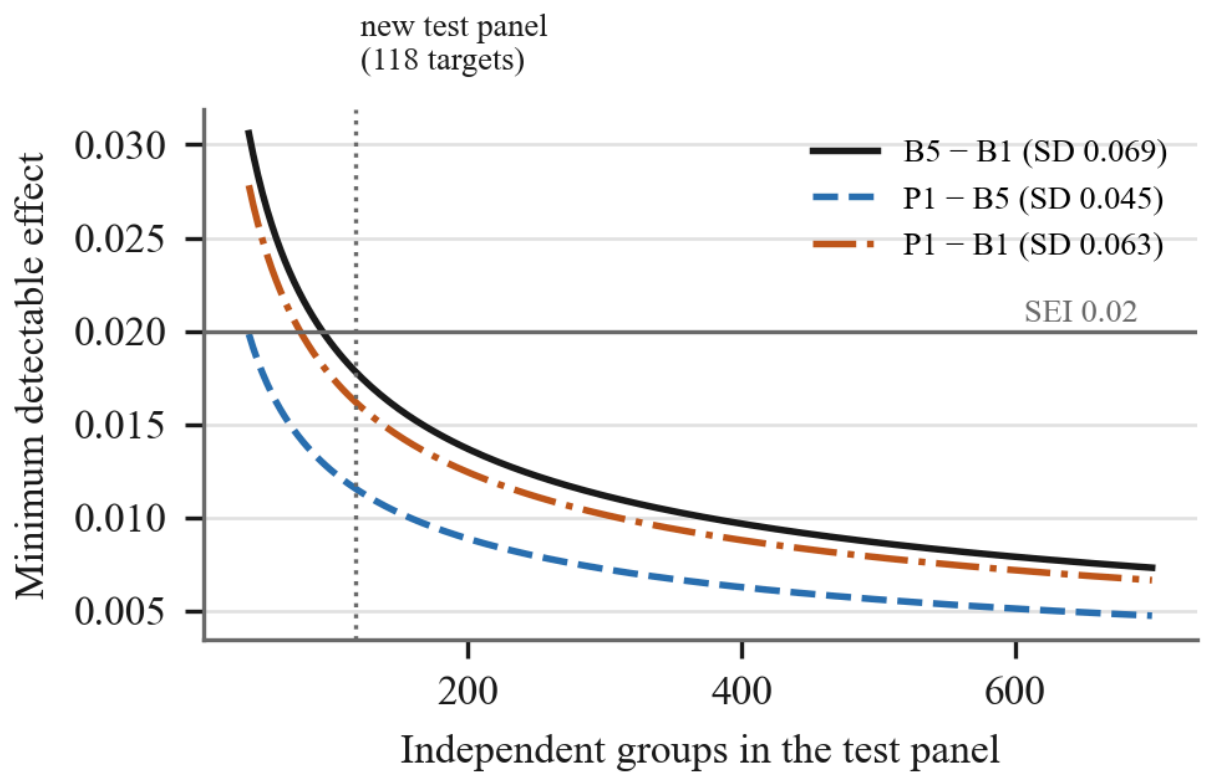

Fig. 12. Minimum detectable effect against the number of independent groups in the test panel.

With 118 target groups the rerun can detect a memory effect of 0.018 and a policy effect of 0.012, both below the 2.0-point threshold. A null result in the rerun is therefore informative: an effect at the threshold would have been found with 80% probability.

### I. Rerun on the Reference-Grouped Split

We reran B1, B5 and P1 on the new split with the same model, temperature, admission thresholds, policy settings and seven-batch training stream (301 training units). The only intended change was the counter-example label in the example-assisted prompt. The rerun made 1,560 distinct model calls (the cache served repeated prompts), used 0.47 million input and 0.37 million output tokens and about 73 minutes of model time. Table XX gives the results with targets as bootstrap groups, next to the model-free baselines on the same units.

TABLE XX. Rerun on the Reference-Grouped Split (374 Units, 118 Target Groups)

| System | Units | RC (95% CI) | Passes |
|---|---|---|---|
| B1 no memory | 374 | 0.149 [0.120, 0.179] | 6 |
| B5 verified memory | 374 | 0.178 [0.124, 0.252] | 8 |
| P1 adaptive agent | 374 | 0.178 [0.124, 0.250] | 8 |
| M1 copy source | 374 | 0.108 [0.085, 0.133] | 0 |
| M2 copy nearest target | 374 | 0.119 [0.104, 0.137] | 4 |
| M3 random training target | 374 | 0.078 [0.068, 0.087] | 1 |

All three LLM systems beat every model-free baseline. B5 and P1 produced identical outputs: the policy never left the spec-first strategy, and no training experience was ever verified, so both systems saw the same counter-examples. Table XXI gives the paired contrasts of (13).

TABLE XXI. Paired Contrasts on the Reference-Grouped Split (Eq. 13, 95% Group-Bootstrap CI)

| Contrast | Groups | Difference (95% CI) |
|---|---|---|
| B5 − B1 | 118 | 0.005 [-0.003, 0.017] |
| P1 − B5 | 118 | 0.000 [0.000, 0.000] |
| P1 − B1 | 118 | 0.005 [-0.003, 0.017] |
| B1 − M2 | 118 | 0.022 [0.005, 0.039] |
| P1 − M2 | 118 | 0.027 [0.008, 0.047] |

Memory gave 0.005 [−0.003, 0.017] over B1 and the policy gave exactly zero over memory. Both intervals sit below the 2.0-point threshold, and the memory interval includes zero. The no-memory baseline beat copying by 0.022 [0.005, 0.039], so the LLM does add agreement that copying cannot reach once the reference path is closed.

Unit-weighted means tell a different story, and the difference matters. B5 averaged 0.178 against 0.149 for B1, a gain of 0.029 per unit. Of that gain, 88% came from a single reference, psql_docs_full_122, which serves 42 of the 374 test units; their RC rose from 0.265 to 0.493. Without this reference the unit-level gain was 0.004, and the group-level contrast was 0.003 [−0.004, 0.014]. Table XXII and Fig. 13 show when this happened.

TABLE XXII. Checkpoints of P1 on the Reference-Grouped Split

| k | Memory | Verified | RC | LG(k) (95% CI) | Churn |
|---|---|---|---|---|---|
| 0 | 0 | 0 | 0.149 | — | — |
| 1 | 50 | 0 | 0.152 | -0.005 [-0.010, 0.000] | 1.00 |
| 2 | 100 | 0 | 0.142 | 0.002 [-0.005, 0.012] | 0.58 |
| 3 | 150 | 0 | 0.177 | 0.006 [-0.003, 0.017] | 0.94 |
| 4 | 200 | 0 | 0.178 | 0.006 [-0.002, 0.017] | 0.06 |
| 5 | 250 | 0 | 0.178 | 0.005 [-0.003, 0.017] | 0.12 |
| 6 | 300 | 0 | 0.178 | 0.005 [-0.003, 0.017] | 0.11 |
| 7 | 301 | 0 | 0.178 | 0.005 [-0.003, 0.017] | 0.00 |

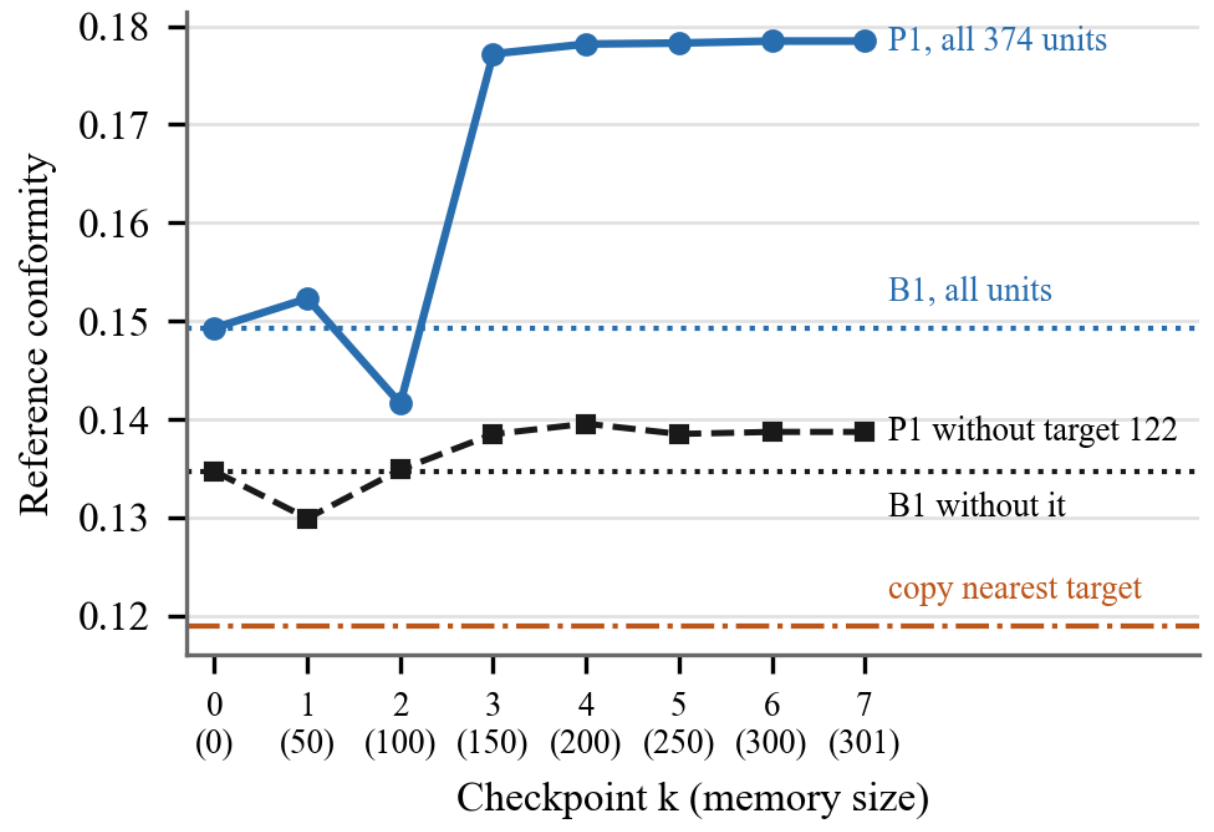

Fig. 13. Reference conformity of P1 on the reference-grouped split, for all units and without the dominant reference (42 units), against B1 and the copy baseline M2.

The jump at k = 3 coincides with a retrieval churn of 0.94: the set of counter-examples changed for almost every unit, and for the 42 units of the dominant reference the new counter-examples pushed the outputs towards its long audit-trigger script. Without that one reference, P1 stays within 0.005 of B1 at every checkpoint. The rerun therefore confirms the audit. Once references cannot cross the split, the memory and policy components show no statistically distinguishable group-level improvement at the level of independent groups, and a unit-weighted learning curve can still look like learning when one large group moves. No experience was verified and no rule was admitted in 301 training units, so the admission and policy mechanisms were never exercised.

# IX. Study 3: Execution-Judged Evaluation

## A. What Accepted, Empty and Judged Outputs Mean

Table XXIII summarises Study 3. Three findings matter for any learning loop. First, a do-nothing agent beat the real agent on the behavioural metric the work first reported. Oracle raises no error on 11 of the first 20 curated units, so an empty script scored 11 of 20 exact matches and a mean fidelity of 0.550, against 1 of 20 and about 0.15 for the agent. On 34 units the do-nothing agent scored 21 of 34.

TABLE XXIII. Execution-Judged Evidence from the Lab Migration Harness

| Measurement | Result |
|---|---|
| Do-nothing vs agent, 20 curated units (exact matches) | 11/20 vs 1/20 |
| Do-nothing vs agent, mean behavioural fidelity | 0.550 vs ≈0.15 |
| Do-nothing, 34 units (exact matches) | 21/34 |
| Accepted verdicts with unresolved inner references | ≈78% |
| Acceptance effects after restricting to resolved verdicts | −6.15 → −0.45 pp; +2.10 → +0.15–0.25 pp |
| Accepted units wrong against Oracle output | 9/93 (10%); 25/114 (22%) |
| LLM-judge verdict churn, two calls, temperature 0 | 16.2% |
| Pre-registered fix cycles FIX-019 to FIX-046 | 13 confirmed, 6 partial, 2 failed, 7 construct-level |
| Demonstrably correct units | 98/216 → 108/216 → 162–170/224 → 167/224 |
| Run-to-run spread, same configuration | ≈10 of 224 units |

Second, acceptance by PostgreSQL proved little: about 78% of accepted verdicts came from PL/pgSQL bodies whose inner references were never resolved. Restricted to resolved verdicts, the two headline acceptance effects shrank from −6.15 to −0.45 points ($p$ = 0.39) and from +2.10 to +0.15–0.25 points ($p$ = 0.33–0.58). Checked against Oracle output, 9 of 93 accepted units (10%) were wrong before a round of fixes and 25 of 114 (22%) after it. Third, an LLM judge used to separate designed from accidental behaviour changed 16.2% of its verdicts between two calls on identical input at temperature 0.

## B. A Gated Fix Cycle as Supervised Admission

The fix cycle is a manual form of the admission step in Section V-C. Each candidate change is one prompt paragraph or one harness fix, tested on the full corpus against predicates written in advance, and kept only if the generated SQL shows the intended effect. Fig. 14 shows the do-nothing comparison and the progress of the cycle.

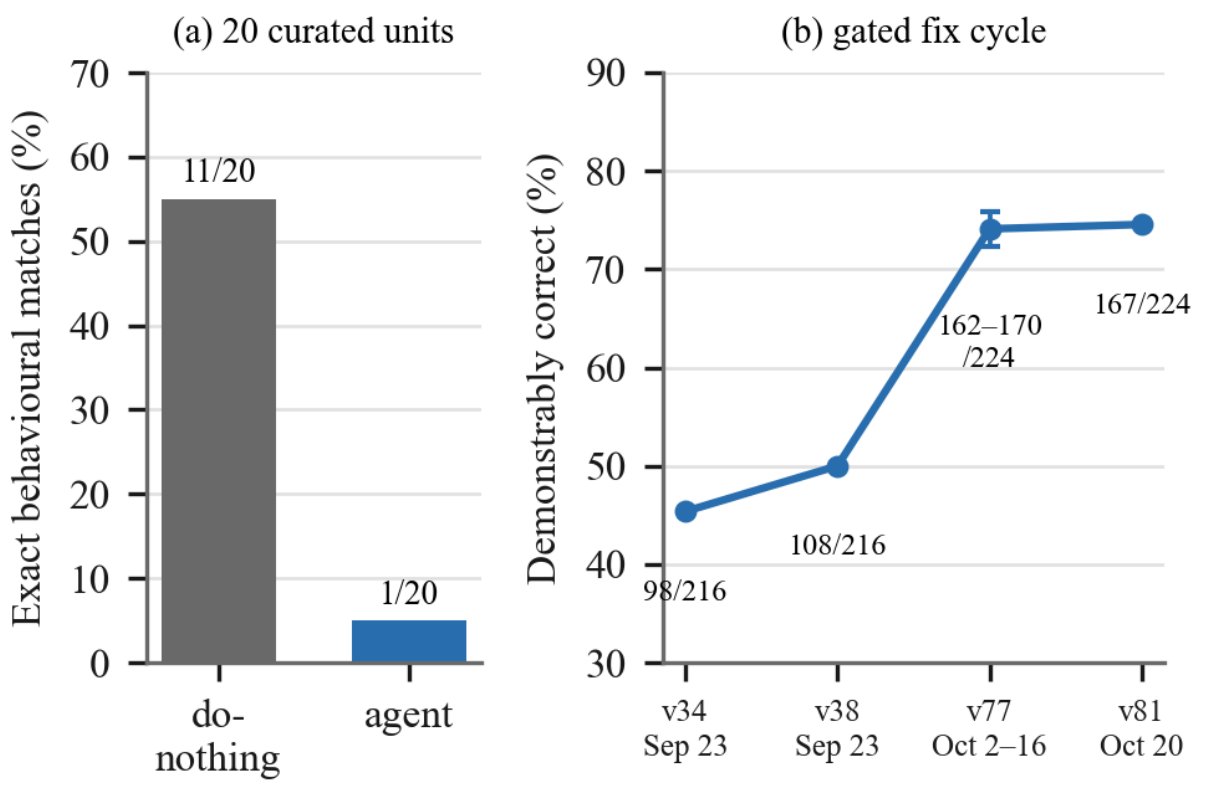

Fig. 14. Study 3: (a) exact behavioural matches of a do-nothing script and the agent on 20 curated units; (b) demonstrably correct units over prompt versions (v77: range of three runs).

Over about four weeks the count of demonstrably correct units rose from 98 of 216 to 167 of 224. Harness fixes explain much of this rise. One resolver defect alone blocked 21 groups, 57% of the blocked cascade, and four harness fixes that cost no tokens raised the migrated units from 53 to 74. Prompt fixes moved the ceiling one construct at a time, for example SELECT INTO STRICT for Oracle's NO_DATA_FOUND, TIMESTAMP(0) for Oracle DATE, cursor expressions, and OUT and IN OUT parameters. Many fixes were partial: a correct fix often exposed the next defect behind it. One fix produced five new accepted units that were all wrong; a whole-file comparison with Oracle output caught them. Two runs of the same configuration differed by about 10 units, so a single run is a point estimate.

# X. Study 4: Knowledge Write-Back at Scale

## A. Margin and Scale

Table XXIV gives the write-back results. In every setting the gap case moved from MISS to HIT right after one correction was ingested, with no re-indexing or restart.

TABLE XXIV. Retrieval of a Written-Back Correction (Top-3, XMLTABLE Gap Case)

| Base size, model | Before | After | Correct | Next best | Margin |
|---|---|---|---|---|---|
| 12, MiniLM | MISS | HIT | 1.004 | 1.567 | 0.563 |
| 120, MiniLM | MISS | HIT | 1.004 | 1.484 | 0.480 |
| 1,200, MiniLM | MISS | HIT | 1.004 | 1.265 | 0.261 |
| 12, BGE-M3-SQL | MISS | HIT | 0.890 | 0.967 | 0.078 |

Fig. 15 plots the margin between the correct rule and the best irrelevant rule. It shrank from 0.563 to 0.261 as the base grew from 12 to 1,200 entries.

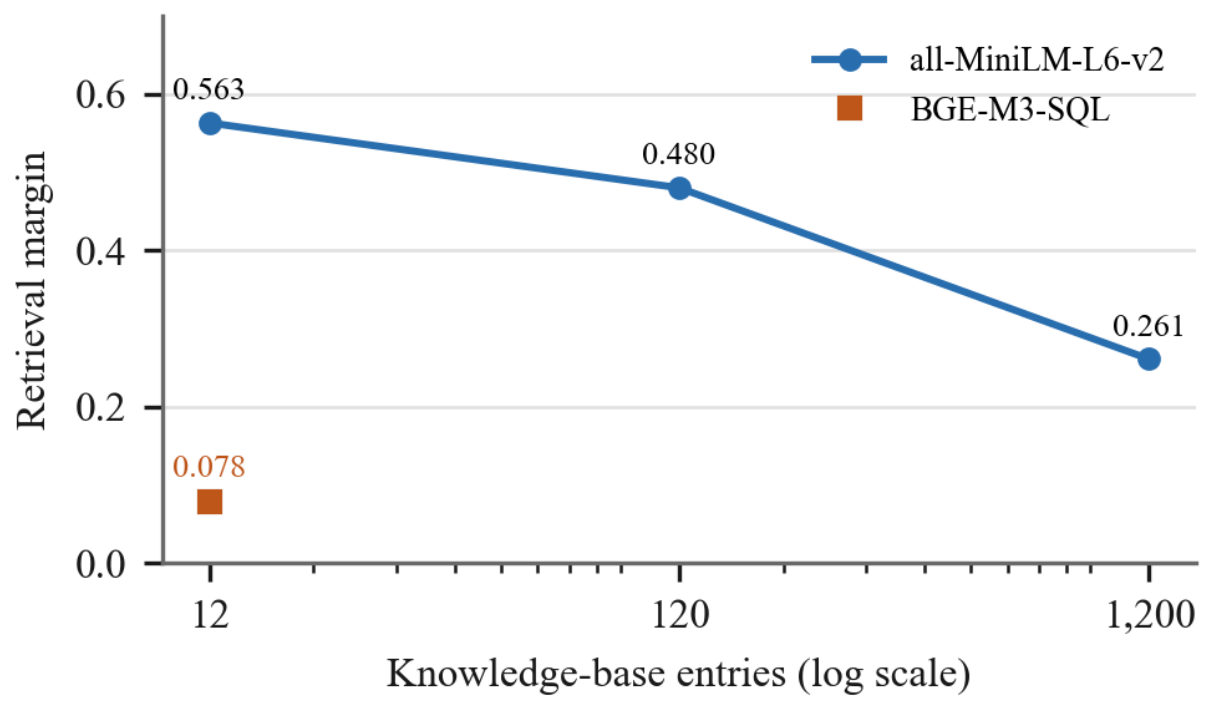

Fig. 15. Retrieval margin of a written-back correction against knowledge-base size, for two embedding models.

The filler rules were synthetic, so this is a trend, not a production estimate. The production embedding model BGE-M3-SQL kept the HIT but cut the margin to 0.078 on the 12-entry base, and embedding 12 documents on a CPU took 242 s against about 1 s for MiniLM.

### B. *Independence of Corrections*

Seven corrections (XMLTABLE, PIVOT, the MODEL clause, BULK COLLECT and FORALL, UTL_FILE, DBMS_OUTPUT and autonomous transactions) were ingested one by one into a 12-entry base. Each became retrievable for its own gap case at once, and none was pushed out of the top three by a later one. Study 4 also explains a pattern in Study 1. When few entries are verified, one entry can sit closest to almost every query. In Study 1 one verified experience was the positive example for 97% of the panel.

## XI. Discussion

### A. *Answers to the Research Questions*

RQ1. Memory raised RC on the full panel, but neither admission nor policy adaptation added anything, and the policy collapsed into one global switch. H1 and H4 were not supported, H2 and H3 could not be tested, and H5 was inconclusive. RQ2. Almost all of the memory gain sat on units that share a reference with training, and copying the nearest training target beat every LLM system. On leak-free units the agent beat copying but did not beat the no-memory baseline. A split grouped by reference closes this path and lowers the copy floor to 0.119. The rerun on it found no memory or policy effect at group level, with enough power to detect 0.018. RQ3. Under execution against Oracle, acceptance and silence-rewarding metrics mislead, and a single LLM judge is unstable. A gated cycle with human verdicts did accumulate gains, mostly by fixing the harness. RQ4. Write-back works at once and without regressions at small scale, but the safety margin shrinks as the base grows and depends on the embedding model.

A related conference presentation by one of the authors [36] reports experiments on five Python repositories. Monitoring adaptation reduced detection lag for all five systems, but improved F1 only for three with lower import coupling and reduced F1 for two with higher coupling. These presentation-level results are not pooled with our migration experiments; they illustrate why averages can conceal negative transfer.

### B. *Rules for Claims of Agent Learning*

The Study 3 harness reached the same conclusion as the non-compensatory rules in [32] by a different route: it reports four separate categories per unit and never a blended score, because each single number it tried rewarded the wrong behaviour. We draw six rules from the four studies:

- Group splits by reference target as well as by source; report how many held-out references occur in training.
- Report a model-free copy baseline next to every learning curve; a gain below copying is not evidence of learning.
- Score a do-nothing agent beside every metric before trusting it.
- Count only verdicts that resolved their references; acceptance by the target system alone is not correctness.
- Repeat LLM-judge calls and report verdict churn; one call at temperature 0 is not deterministic.
- Report the state behind the curve: verified entries, distinct examples shown, policy entropy and admitted rules.

Applied to Study 1, the rules turn a positive-looking result into a null result for learning. Applied to Study 3, they make the reported gains believable, because each gain is tied to a pre-registered change and read against Oracle output. The first two rules cost no model calls, and Section VIII shows that both can be applied to a finished run.

## XII. Threats to Validity

Construct validity is the main threat. In Studies 1 and 2, RC measures agreement with a topically related documentation example, not with a translation, and no pair was reviewed. Internal validity: Study 1 is a single run per system at temperature 0, ablations were not retrained, and the training stream used 311 of 1,099 units. The Stage A audit and the dependency graph were not available, so two parts of the design were not tested. On the reference-grouped split only B1, B5 and P1 were rerun; B2, B4, the ablations and the controls were not. External validity: Study 1 uses one documentation corpus and one LLM; Study 3 uses one harness, one model routing and 34 files. Studies 3 and 4 are reported from the lab's records; their raw artifacts are internal. Table XXV lists the threats and their status.

TABLE XXV. Validity Threats and Their Status

| Threat | Status in this study |
|---|---|
| Reference leakage through shared targets | Observed: 264 of 367 test units; audited; removed in the new split |
| Weak golden pairs | Observed: topical references, 0 reviewed pairs |
| Static versus behavioural correctness | Observed in Study 3: 10–22% of accepted units wrong |
| Degenerate metric optima | Observed in Study 3; RC gives empty output zero |
| Low success rate | Observed: ≤ 8 passes per system; NTR and TPS unstable |
| Untested design parts | Specification audit and dependency graph not available |
| Run-to-run variance | Not estimated in Study 1; ≈10/224 units in Study 3 |
| Generalisation | One corpus and model (Study 1); one harness (Study 3) |

Two threats are partly mitigated by Study 2: reference leakage is removed in the new split, and the copy and do-nothing baselines bound what RC can reward. The others remain open and shape the future work below.

## Conclusions

We built an evidence-graded self-learning agent for Oracle-to-PostgreSQL migration and tested whether it learns. On a documentation-derived corpus, memory raised reference conformity by 0.015, admission and policy adaptation added nothing, and the policy collapsed to one strategy. An audit showed that 72% of held-out units shared their reference with training. On leak-free units the memory gain disappeared, and copying the nearest training target beat every LLM system on the full panel. The answer to the title question is therefore no for this corpus.

A split grouped by reference removes the leak and lowers the copy floor from 0.222 to 0.119. A rerun on it found no memory effect (0.005 [−0.003, 0.017]) and no policy effect at the level of 118 independent reference groups; the unit-weighted gain of 0.029 came almost entirely from one reference shared by 42 units. The lab studies show what a sound learning signal needs. Judged by execution against Oracle, a pre-registered, human-gated fix cycle raised demonstrably correct units from 98 of 216 to 167 of 224, while acceptance, silence and single judge calls would each have misled an automatic loop. Write-back of corrections works but loses margin as the knowledge base grows.

Future work will review a stratified sample of pairs, rerun the remaining systems on the reference-grouped split, run the agent with per-unit source specifications, and replace the reference-conformity signal with execution evidence from the Study 3 harness, so that an automatic proposer can draft fixes while the gate keeps the verdicts.

## Declaration on Generative AI

During the preparation of this work, the authors used ChatGPT in order to: Grammar and spelling check. After using this service, the authors reviewed and edited the content as needed and take full responsibility for the publication's content.

## Funding and Data Availability

Funding: this work was carried out within the EPAM Systems GenAI R&D Lab and received no external funding.

Data availability: the Study 1 corpus is derived from public Oracle and PostgreSQL documentation. Manifests, splits (original and reference-grouped), run logs, snapshots and the Study 2 scripts are available from the authors on reasonable request. The Study 3 harness and corpus and the Study 4 code are internal to the lab.